\documentclass[journal=jctcce,manuscript=article]{achemso}
\usepackage{amssymb,amsmath,amsfonts,multicol,multirow,longtable,array,mathpazo}
\usepackage{lscape}
\usepackage[version=3]{mhchem} 
\usepackage{appendix}
\usepackage{soul} 
\usepackage[usenames]{color}
\usepackage{makecell}
\usepackage{enumerate}
\usepackage{xr}
\usepackage[T1]{fontenc} 
\usepackage{booktabs}
\usepackage{tikz}
\usetikzlibrary{shapes.geometric}
\usepackage{threeparttable}
\usepackage{graphicx}
\usepackage{subfigure}
\usepackage{colortbl}
\usepackage{framed}
\usepackage{verbatim}
\usepackage{amsbsy}
\usepackage{bm}
\usepackage{bbm}
\usepackage[normalem]{ulem}
\usepackage{float}
\usepackage[linesnumbered,boxed]{algorithm2e}
\usepackage{algorithm2e}
\usepackage{float}
\usepackage{subfigure}
\usepackage{simplewick} 

\newcounter{xscheme}

\newfloat{Algorithm}{htbp}{alg}
\floatname{Algorithm}{Algorithm}

\newcounter{exe}[figure]
\newcommand{\iexe}{\refstepcounter{exe}\the\value{exe}:}

\setkeys{acs}{maxauthors = 0} 

\author{Chunzhang Liu}
\author{Ning Zhang}
\author{Wenjian Liu}\email{liuwj@sdu.edu.cn}
\affiliation{Qingdao Institute for Theoretical and Computational Sciences and Center for Optics Research and Engineering,
	Shandong University, Qingdao, Shandong 266237, China}

\title{Putting PASPT2 on a Firmer Basis}

\setkeys{acs}{articletitle=true}

\begin{document}
	
\begin{abstract}
The recently proposed partial-active-space (PAS) multi-state second-order perturbation theory (PASPT2)
[Precis. Chem. 4, 997 (2026)] features connected amplitudes and a connected and
closed intermediate Hamiltonian. Despite these hallmarks, PASPT2 (denoted as PASPT2H from now on)
is not strictly size-extensive, as originally thought (and numerically confirmed), albeit strictly size-consistent.
Nevertheless, PASPT2 is near-extensive for the states with major
projections on the chosen PAS $\mathcal{M}_0$. This becomes more transparent upon introducing PASPT2X,
a strictly size-extensive variant. Compared with the intruder-prone PASPT2X, the intruder-free PASPT2H merely neglects second-order corrections
that are important only for those states with major projections on the orthogonal complement $\mathcal{R}_X$ of $\mathcal{M}_0$ within
the closed space $\mathcal{M}_X$ ($=\mathcal{M}_0\oplus\mathcal{R}_X$). Such states are usually not supported
by the chosen finite one-particle basis set. The weak violation of size-extensivity
is therefore numerically insignificant for the target states supported by $\mathcal{M}_0$, reinforcing the theoretical basis of PASPT2H.
\end{abstract}

\noindent
Keywords: strong correlation, partial active space, closed operator, quasi-open operator, open operators, connectivity of cluster amplitudes,
connectivity and closure of effective Hamiltonian, size-extensive, near-extensive

\maketitle

\clearpage
\newpage
\section{Introduction}
Strongly correlated systems, especially those with many unpaired electrons, feature both strong static and strong dynamic correlations.
The former stems from strong couplings among near-degenerate configurations and demands
a multi-configurational description even just for a qualitative description, whereas the latter
originates from contributions of high-lying configurations characterized by small
coefficients in the target wave function. The major challenge of such systems is that the static and dynamic components of correlation
are strongly entangled and even interchangeable, thereby rendering a preset partitioning of the two generally invalid.
To guide future development of \emph{ab initio} quantum chemical methods for accurately describing such systems,
we introduce the following criteria for a good wave function Ansatz:
\begin{enumerate}[(1)]
\item Adaptive treatment of static and dynamic correlations.
This is necessitated by the strong entanglement and mutual interchangeability of the two components, which renders
any structured reference space (e.g., complete active space (CAS) or generalized valence bond) inherently inflexible and unbalanced.
The inflexibility of a preselected structured reference space lies in that it is not of the same quality
for all target states at all geometries.
This issue can largely be alleviated by choosing a sufficiently large CAS. However, a large CAS is itself problematic in many aspects\cite{iCAS}: for instance,
how to select which and how many orbitals as active orbitals, how to maintain a consistent CAS during orbital optimization
and along potential energy surface scans, not to mention the combinatorial explosion in computational cost.
Moreover, a large CAS usually contains many configurations that are
energetically higher than many configurations in the external space. Treating the former exactly but the latter approximately is inherently imbalanced.
All these pinpoint to the use of a general model space (GMS) devoid of any special structure, which
can readily be generated by selected configuration interaction (sCI).

\item  Size-extensive. Given a GMS, a method that is size-extensive is still highly desired. Note that
size-extensivity (or connectedness of the energy)\cite{Size-extensivity,nooijen2005reflections} is a mathematical property of the method itself.
It means that the calculated energy scales asymptotically linearly with respect to the number of interacting electrons, which
is a necessary condition for uniform accuracy as the system size increases--only size-extensive methods can be applied
to extended systems.

\item Size-consistent. Given a GMS, a size-consistent method is also highly desired. Different from size-extensivity, which is independent of
physical systems, size-consistency\cite{Size-consistency} refers to the additive separability of the energies
of a composite system AB composed of non-interacting subsystems A and B. Specifically, when the GMS of AB
is taken as the direct product of those of A and B, the energies $\{E_{AB}(ij)\}_{(ij)=1}^{M_AM_B}$ calculated
for the whole AB system
should be precisely $\{E_{\mathrm{A}}(i)+E_{\mathrm{B}}(j)\}_{(ij)=1}^{M_AM_B}$,
with $\{ E_{\mathrm{A}}(i)\}_{i=1}^{M_{\mathrm{A}}}$ and
$\{E_{\mathrm{B}}(j)\}_{j=1}^{M_{\mathrm{B}}}$ being the respective energies of A and B calculated separately.
A size-consistent method need not be size-extensive. Sometimes a theoretically size-inconsistent method can be made numerically size-consistent\cite{PASPT2}.

\item Intruder-free. Given a GMS, the intruder-state problem\cite{Intruder1972} that plagues nonvariational methods can persist.
In particular, methods that are immune to this problem for low-lying states may encounter it again for high-lying states\cite{NEVPT2Intruder1,NEVPT2Intruder2}.
Instead of using empirical remedies\cite{CASPT2Intruder,CASPT2IntruderImaginary}, an intrinsically intruder-free method is highly desired.

\item  Invariant under orbital rotations. Ideally, computed energies are independent of the choice of molecular orbitals.
Since different orbital sets are related by unitary transformations (or orbital rotations),
it is highly desirable for the method to be invariant under such transformations within each orbital class\cite{OrbitalInvariance}.

\item  Easily spin adapted for arbitrary open-shells. Without explicit spin adaptation, it is exceedingly difficult--if not impossible--for
any method to describe correctly low-spin states in systems with many unpaired electrons.

\item  Simultaneous treatment of multiple states. While one-state-at-a-time methods remain useful for certain applications,
they suffer from inherent limitations when dealing with situations where multiple states are nearly degenerate and even crossed. Therefore,
the ability of treating simultaneously multiple states--especially in open-shell systems--is highly desirable.

\item Continuous potential energy surfaces. This is usually achieved by using a consistent CAS across geometries.
In contrast, an automatically generated GMS (e.g., via sCI) may vary along geometric coordinates.
Nevertheless, it is still possible for a GMS-based method to yield continuous potential energy surfaces in two ways.
One is to stick to the same set of configurations across geometries, and the other is to retain
all configurations with appreciable coefficients in the GMS, so that  the coefficients
of the outer configurations can be estimated by perturbation with sufficient accuracy
relative to those by diagonalization. The former is, of course, not of the same accuracy at different geometries.

\item  Near-exact. If desired, the method can be made to approach full configuration interaction (FCI) systematically, by enlarging the GMS.

\item Readily combined with relativity and even quantum electrodynamics (QED). The ultimate goal of electronic structure calculations
is to make the left- and right-hand sides of the equation ``relativity + correlation + QED = experiment'' as close as possible\cite{LiuWIRES2023}.
Therefore, the method should readily be combined with relativistic-QED Hamiltonians\cite{eQED,LiuPerspective2020} or
spin-separated exact two-component (X2C) relativistic Hamiltonians\cite{X2CSOC1,X2CSOC2} for treating spin-orbit coupling and correlation on an equal footing.
They require the use of spin-dependent and spin-free configuration state functions as the many-electron basis, respectively.
\end{enumerate}

It is evident that no approximate wave function method can satisfy simultaneously all of the above criteria. Therefore,
what one can do is to formulate a method that fulfills as many of them as possible.
Motivated by the desire of having an adaptive and size-extensive many-body theory for multiple states of strongly correlated
electrons, we have recently developed a partial-active-space (PAS) second-order perturbation theory (PASPT2)\cite{PASPT2}.
Here, the term PAS is regarded as equivalent to GMS\cite{GMSPT2a,GMSPT2b,GMSPT2c}
(also called incomplete model space (IMS) or incomplete active space (IAS) in the literature)
that emphasizes the complete lack of structure. In contrast, other terms (e.g.,
general active space (GAS)\cite{GAS1,GAS2}, SplitGAS\cite{SplitGas1,SplitGas2}, restricted active space (RAS)\cite{RAS1,RAS2,RAS3},
and occupation-restricted multiple active space (ORMAS)\cite{ORMAS}, etc.) all have certain structures, albeit more flexible than CAS.
In particular, the PAS need not be a subspace of CAS. Rather, it can be that
generated by sCI\cite{iCIPT2,iCIPT2New} over the full Hilbert space. Anyway,
the emphasis here is size-extensivity even for an unstructured model space.
It turns out that PASPT2 satisfies all the above criteria except for the energy invariance under rotation of active orbitals,
which is inherited from the Jeziorski-Monkhorst (JM) Ansatz\cite{JM1981} for the wave operator.
As a matter of fact, all methods rooted in this Ansatz lack orbital invariance, even when a CAS is
taken as the reference space\cite{OrbitalInvariance,SS-MRPT2a,SS-MRPT2c}. To what extent the lack of orbital invariance
affects the PASPT2 results remains to be examined systematically. Here, we focus on one aspect of PASPT2,
the size-extensivity problem. While the size-extensivity of single-reference methods can readily be checked either algebraically or diagrammatically,
it is not the case for multi-reference methods, especially when a diagonalizaton step has to be taken to obtain the energies, as done in PASPT2.
It appears that the connectedness of the amplitudes plus the connectedness and closure of the Hamiltonian
are often not sufficient for the energies to be connected. Rather,
 the Hamiltonian matrix should additionally exhibit a specific structure\cite{Mukerjee1989M0,Bartlett1990M0,PASPT2Eigen} (vide post).
It will be shown that PASPT2 is not strictly size-extensive, as originally thought (and numerically confirmed), albeit strictly size-consistent.
However, it is near-extensive and the weak violation of size-extensivity has hardly numerical consequence on the target states.
To this end, we will derive two new variants (PASPT2M and PASPT2X) of PASPT2 and demonstrate that our original variant (to be renamed to PASPT2H)
is indeed most recommended for practical use.

The following conventions are to be used. The occupied and unoccupied spin orbitals in the chosen Fermi vacuum $|\alpha\rangle$
are denoted by $\{I, J,\cdots\}$ and $\{A, B, \cdots \}$, respectively.
The former are further split into inactive holes $\{i,j,\cdots\}$ and active holes $\{v, v_1, v_2,\cdots \}$, whereas the latter
are further split into active particles $\{u, u_1, u_2,\cdots\}$ and inactive particles $\{a, b, \cdots\}$.
The active holes and particles are also called collectively ``local active spin orbitals'' (LASO)\cite{GMS-SU-CC2003}
that distinguish $|\alpha\rangle$ from another reference determinant (DET) $|\beta\rangle$, whereas the inactive holes and particles
are common to $|\alpha\rangle$ and $|\beta\rangle$. General spin orbitals are denoted as $\{p,q, \cdots\}$.
The symbol $\{O\}_\alpha$ denotes operator $O$ normal-ordered with respect to $|\alpha\rangle$.
The Einstein convention over repeated indices is always employed.




\section{PASPT2}
\subsection{Second-order Effective Hamiltonian}
The starting point is the JM multi-exponential Ansatz\cite{JM1981} for the Hilbert-space or state-universal (SU) wave operator $\Omega$,
\begin{align}
\Omega&=\sum_{\alpha\in\mathcal{M}_0}e^{T^\alpha}P_{\alpha}=\Omega P,\label{Omega}\\
P&=\sum_{\alpha\in\mathcal{M}_0}P_{\alpha},\quad P_{\alpha}=|\alpha\rangle\langle\alpha|,
\end{align}
where $P$ is the projector onto the model $\mathcal{M}_0=\{|\alpha\rangle, |\beta\rangle, |\gamma\rangle,\cdots\}$.
Eq. \eqref{Omega} implies that
\begin{align}
\Omega Q &=0,\quad Q= 1-P, \label{OmegaQ}\\
\Omega&=\Omega^{(0)}+\sum_{i=1}\Omega^{(i)}=P+\sum_{i=1}\Omega^{(i)},
\end{align}
where the superscript $(i)$ denote the ordering of $\Omega$. In addition to these,
the intermediate normalization (IN)
\begin{align}
P\Omega P&=P\Omega^{(0)} P=P\label{POP}
\end{align}
has been invoked to formulate\cite{JM1981}
multi-reference coupled-cluster theory (MRCC) for a CAS $\mathcal{M}_C$ (instead of
the present PAS $\mathcal{M}_0$). The IN implies that the wave operator does not couple different reference functions, viz.,
\begin{align}
\langle \beta|\Omega|\alpha\rangle&=\langle \beta|e^{T^\alpha}|\alpha\rangle=\delta_{\beta\alpha},\label{INC}
\end{align}
which further implies that all internal cluster operators (which only scatter among reference functions)
can be set to zero without introducing approximations for a CAS.
Realizing that this is no longer the case for a PAS $\mathcal{M}_0$, Li and Paldus\cite{GMS-SU-CC2003}
introduced the so-called connectivity condition (or C-condition) to determine the amplitudes
of internal clusters. For instance, if the operator $a^u_v/a^{u_1u_2}_{v_1v_2}$ generates a reference DET
when operating on $|\alpha\rangle$, the C-condition reads
\begin{subequations}\label{C-Cond}
\begin{equation}
C^\alpha_{u,v}=t^\alpha_{u,v}=0,\label{1BCond}
\end{equation}
\begin{equation}
C^\alpha_{u_1v_1,u_2v_2}=t^\alpha_{u_1v_1, u_2v_2}+t^\alpha_{u_1,v_1} t^\alpha_{u_2,v_2}-t^\alpha_{u_2,v_1} t^\alpha_{u_1, v_2}=0.\label{2BCond}
\end{equation}
\end{subequations}
Eq. \eqref{2BCond} determines $t^\alpha_{u_1v_1, u_2v_2}$ when $\{a_{v_1}^{u_1}\}_\alpha$ and $\{a_{v_2}^{u_2}\}_\alpha$
and/or $\{a_{v_1}^{u_2}\}_\alpha $ and $\{a_{v_2}^{u_1}\}_\alpha$
are external single excitations from $|\alpha\rangle$. They further showed\cite{GMS-SU-CC2003,PaldusJMC2004} that
the resulting IN-GMS-SU-CC theory is size-extensive. However, this declaration is disapproved during the course
of formulating PASPT2\cite{PASPT2} by perturbative expansion of IN-GMS-SU-CCSD (coupled-cluster with singles and doubles):
there exists a disconnected term in the amplitude equation,
which can only be removed to first order (by introducing a special zeroth-order Hamiltonian; vide post),
but shows up again beyond the first order. This finding reconfirms the statement\cite{MRconnectivity} that the IN \eqref{POP} must be abandoned
in order to have a GMS-based size-extensive theory; instead, a general normalization (GN)
determined by the cluster operators themselves must be adopted.

Inserting the wave operator \eqref{Omega} into the Schr\"odinger equation leads to the generalized Bloch equation\cite{Bloch1958,LindgrenBook}
 \begin{align}
e^{-T^\alpha} H e^{T^\alpha} |\alpha\rangle
&= \sum_{\gamma\in \mathcal{M}_0} e^{-T^\alpha} e^{T^\gamma}|\gamma\rangle  H^{\mathrm{eff}}_{\gamma\alpha}, \label{GenBlochT}\\
H^{\mathrm{eff}}_{\gamma\alpha}&=\langle\gamma| H^{\mathrm{eff}} |\alpha\rangle.
\end{align}
The effective Hamiltonian $ H^{\mathrm{eff}}$ is determined by the condition (without assuming the IN)
 \begin{align}
\langle\beta|e^{-T^\alpha} H e^{T^\alpha} |\alpha\rangle
&= \sum_{\gamma\in \mathcal{M}_0}\langle\beta| e^{-T^\alpha} e^{T^\gamma}|\gamma\rangle  H^{\mathrm{eff}}_{\gamma\alpha}, \label{HeffCond}
\end{align}
which can be expanded as
\begin{align}
&\langle\beta|H+[H,T^\alpha]+\frac{1}{2}[[H, T^\alpha],T^\alpha]+\cdots|\alpha\rangle\nonumber\\
&=\sum_{\gamma\in \mathcal{M}_0}\langle\beta|1+(T^\gamma-T^\alpha)+\frac{1}{2}[T^\gamma,T^\alpha]+\frac{1}{2}(T^\gamma-T^\alpha)^2+\cdots|\gamma\rangle H_{\gamma\alpha}^{\mathrm{eff}}\nonumber\\
&=H_{\beta\alpha}^{\mathrm{eff}}+\sum_{\gamma\in \mathcal{M}_0, \gamma\ne\alpha}\langle\beta|(T^\gamma-T^\alpha)+\frac{1}{2}[T^\gamma,T^\alpha]+\frac{1}{2}(T^\gamma-T^\alpha)^2+\cdots|\gamma\rangle H_{\gamma\alpha}^{\mathrm{eff}}.
\end{align}
By partitioning the Hamiltonian and expanding the cluster operators $T^\alpha$ as
\begin{align}
       H&=H_0^\alpha+V^\alpha,\\
T^\alpha&=\sum_{i=1}^\infty T^{\alpha(i)},
\end{align}
we obtain
\begin{align}
H_{\beta\alpha}^{\mathrm{eff}(0)}&=\langle\beta|H_0^\alpha|\alpha\rangle,\\
H_{\beta\alpha}^{\mathrm{eff}(1)}&=\langle\beta|V^\alpha+ [H_0^\alpha, T^{\alpha(1)}]|\alpha\rangle
-\sum_{\gamma\in \mathcal{M}_0, \gamma\ne\alpha}\langle\beta|(T^{\gamma(1)}-T^{\alpha(1)})|\gamma\rangle H_{\gamma\alpha}^{\mathrm{eff}(0)},\label{Heff1a}\\
H_{\beta\alpha}^{\mathrm{eff}(2)}&=\langle\beta|[V^\alpha, T^{\alpha(1)}]+[H_0^\alpha, T^{\alpha(2)}]+\frac{1}{2}[[H_0^\alpha, T^{\alpha(1)}],T^{\alpha(1)}]|\alpha\rangle \nonumber\\
&-\sum_{\gamma\in \mathcal{M}_0, \gamma\ne\alpha}\langle\beta|(T^{\gamma(1)}-T^{\alpha(1)})|\gamma\rangle H_{\gamma\alpha}^{\mathrm{eff}(1)}\nonumber\\
&-\sum_{\gamma\in \mathcal{M}_0, \gamma\ne\alpha}\langle\beta|(T^{\gamma(2)}-T^{\alpha(2)})+\frac{1}{2}[T^{\gamma(1)},T^{\alpha(1)}]+\frac{1}{2}(T^{\gamma(1)}-T^{\alpha(1)})^2|\gamma\rangle H_{\gamma\alpha}^{\mathrm{eff}(0)}.
\end{align}
It is clear that both $H_{\gamma\alpha}^{\mathrm{eff}(1)}$ and $H_{\gamma\alpha}^{\mathrm{eff}(2)}$ have to be simplified for practical use.
The simplest option is
\begin{subequations}\label{H0Cond}
\begin{equation}
H_{\beta\alpha}^{\mathrm{eff}(0)}=\langle\beta|H_0^\alpha|\alpha\rangle=E_\alpha^{(0)}\delta_{\alpha\beta},\quad \alpha,\beta\in\mathcal{M}_0,\label{Cond0}
\end{equation}
\begin{equation}
\langle\beta|[H_0^\alpha, T^{\alpha(i)}]|\alpha\rangle=0,\quad i=1, 2,\label{CondT0}
\end{equation}
\begin{equation}
\langle\beta|[[H_0^\alpha, T^{\alpha(1)}], T^{\alpha(1)}]|\alpha\rangle=0,\label{CondT1}
\end{equation}
\end{subequations}
in terms of which $H_{\gamma\alpha}^{\mathrm{eff}(1)}$ and $H_{\gamma\alpha}^{\mathrm{eff}(2)}$ are simplified to
\begin{align}
H_{\beta\alpha}^{\mathrm{eff}(1)}&=\langle\beta|V^\alpha|\alpha\rangle,\label{Veff1}\\
H_{\beta\alpha}^{\mathrm{eff}(2)}&=\langle\beta|[V^\alpha, T^{\alpha(1)}]|\alpha\rangle
-\sum_{\gamma\in \mathcal{M}_0, \gamma\ne\alpha}\langle\beta|(T^{\gamma(1)}-T^{\alpha(1)})|\gamma\rangle H_{\gamma\alpha}^{\mathrm{eff}(1)}\label{Veff2a}\\
&=\langle\beta|[H, T^{\alpha(1)}]|\alpha\rangle
-\sum_{\gamma\in \mathcal{M}_0, \gamma\ne\alpha}\langle\beta|(T^{\gamma(1)}-T^{\alpha(1)})|\gamma\rangle H_{\gamma\alpha}. \label{Veff2}
\end{align}
The second (renormalization) term on the right-hand side of Eq. \eqref{Veff2a} or \eqref{Veff2} vanishes identically
for a closed model space $\mathcal{M}_X$ [cf. Eq. \eqref{CondTotalMx}], but is generally
disconnected in the case of PAS $\mathcal{M}_0$: for instance, when
$\langle\beta|T^{\alpha(1)}|\gamma\rangle=t^{\alpha(1)}_{\beta,\gamma}$ but $\langle\beta|T^{\gamma(1)}|\gamma\rangle=0$ [cf. Eq. \eqref{CondT}],
the former may not have a common orbital index with the matrix element $H_{\gamma\alpha}^{\mathrm{eff}(1)}$ or $H_{\gamma\alpha}$, for the relation
between $|\beta\rangle$ and $|\gamma\rangle$ dictated by $T^{\alpha(1)}$ is not bound to that between $|\gamma\rangle$ and $|\alpha\rangle$
in $\mathcal{M}_0$. This term should therefore be ignored,  thereby leading to the following effective Hamiltonian correct to second order\cite{PASPT2}
\begin{align}
H^{\mathrm{eff}[2]}&=P(H+\sum_{\alpha\in\mathcal{M}_0}[H, T^{\alpha(1)}]P_\alpha) P, \label{HeffFinal}
\end{align}
which is manifestly connected. The diagonal elements are simply the energies
of the corresponding reference functions by (non-Hartree-Fock) second-order M{\o}ller-Plesset perturbation theory (MP2)\cite{MP2}, viz.
\begin{align}
E_{\alpha}^{[2]}=E_{HF}(\alpha)+E^{(2)}_\alpha.\label{MP2Ene}
\end{align}

Eq. \eqref{H0Cond} imposes restrictions on the zero-order Hamiltonian $H_0^\alpha$. However, they are not enough to determine
a unique form for  $H_0^\alpha$. To search for additional constraints on $H_0^\alpha$, we spell out  Eq. \eqref{CondT0}
\begin{align}
\langle\beta|[H_0^\alpha, T^{\alpha(i)}]|\alpha\rangle&=\langle\beta|H_0^\alpha T^{\alpha(i)}|\alpha\rangle-\langle\beta|T^{\alpha(i)} H_0^\alpha|\alpha\rangle\nonumber\\
&=\sum_{\delta\in P+Q} \langle\beta|H_0^\alpha|\delta\rangle\langle\delta| T^{\alpha(i)}|\alpha\rangle
-\sum_{\delta\in P+Q} \langle\beta|T^{\alpha(i)}|\delta\rangle\langle\delta|H_0^\alpha |\alpha\rangle \nonumber\\
&=\sum_{\delta\in P+Q} \langle\beta|H_0^\alpha|\delta\rangle\langle\delta| T^{\alpha(i)}|\alpha\rangle-  \langle\beta|T^{\alpha(i)}|\alpha\rangle E_\alpha^{(0)}.\label{temp1}
\end{align}
Use of the conditions \eqref{OmegaQ} and \eqref{Cond0} has been made to arrive at Eq. \eqref{temp1}.
The first attempt to further simplify Eq. \eqref{temp1} is to invoke the T-condition
\begin{align}
PT^{\alpha(i)}|\alpha\rangle&=0,\quad i=1,2,\quad \alpha\in\mathcal{M}_0, \label{CondT}
\end{align}
so as to obtain
\begin{align}
\langle\beta|[H_0^\alpha, T^{\alpha(i)}]|\alpha\rangle
&=\sum_{\delta\in Q} \langle\beta|H_0^\alpha|\delta\rangle\langle\delta| T^{\alpha(i)}|\alpha\rangle.\label{temp2}
\end{align}
However, the matrix elements $\langle\beta|H_0^\alpha|\delta\rangle$ for $\beta\in P$ and $\delta\in Q$ still remain undefined.
The alternative try to simplify Eq. \eqref{temp1} is to require
\begin{align}
\langle\beta|H_0^\alpha|\delta\rangle=E_{\delta}^{(0)}\delta_{\beta\delta},\quad \beta\in P,\quad \delta\in P+Q \label{DiagH0}
\end{align}
from the outset, so as to obtain
\begin{align}
\langle\beta|[H_0^\alpha, T^{\alpha(i)}]|\alpha\rangle
&=(E_\beta^{(0)}-E_\alpha^{(0)}) \langle\beta|T^{\alpha(i)}|\alpha\rangle=0,\quad i=1,2, \label{Tcond}
\end{align}
which leads naturally to the T-condition \eqref{CondT} (since $E_\beta^{(0)}\ne E_\alpha^{(0)}$ in general). Eq. \eqref{DiagH0} dictates that $H_0^\alpha$
of the same for the $P$ and $Q$ spaces
should take all DETs, whether internal or external, as its eigenfunctions, viz.
\begin{align}
H_0^\alpha|\delta\rangle=E_\delta^{(0)}|\delta\rangle,\quad \delta\in P+Q,\label{Cond00}
\end{align}
which can only be fulfilled by the following reference-specific but orbital-diagonal form
\begin{align}
H_0^\alpha=\epsilon_p(\alpha) a^p_p=\epsilon_p(\alpha) n_p(\alpha)+\epsilon_p(\alpha) \{a^p_p\}_\alpha,\label{Cond00OP}
\end{align}
where $\epsilon_p(\alpha)=f_p^p(\alpha)$, with $f_p^q(\alpha)$ being the Fock matrix elements
specific to $|\alpha\rangle$ [cf. Eq. \eqref{Fspin}]. We then have
\begin{align}
P[[[H_0^\alpha, T^\alpha], T^\alpha] \cdots ]P_\alpha=0.
\end{align}
Literally, all commutators between $H_0^\alpha$ and $T^\alpha$ have no nonzero matrix elements over the reference functions,
which is another desired property of $H_0^\alpha$.

In summary, the desired forms for the first- and second-order effective Hamiltonians in
Eqs. \eqref{Veff1} and \eqref{Veff2}, respectively, arise only under the conditions \eqref{Cond00OP} and \eqref{CondT}.
Note in passing that the first-order T-condition
in Eq. \eqref{CondT} is just the first-order C-condition (cf. Eq. \eqref{C-Cond}). However, the second-order T-condition therein is different from
the second-order C-condition for, e.g., the second-order amplitude $t^{\alpha(2)}_{u_1v_1, u_2v_2}$ of the
internal double excitation $\{a^{u_1u_2}_{v_1v_2}\}_\alpha$ from $|\alpha\rangle$ (cf. Eq. \eqref{2BCond})
\begin{align}
t^{\alpha(2)}_{u_1v_1, u_2v_2}&=-t^{\alpha(1)}_{u_1,v_1} t^{\alpha(1)}_{u_2,v_2}+t^{\alpha(1)}_{u_1,v_2} t^{\alpha(1)}_{u_2,v_1},
\end{align}
which is nonzero when $\{a_{v_1}^{u_1}\}_\alpha$ and $\{a_{v_2}^{u_2}\}_\alpha$ and/or $\{a_{v_2}^{u_1}\}_\alpha$ and $\{a_{v_1}^{u_2}\}_\alpha$ are external
single excitations from $|\alpha\rangle$. Eq. \eqref{CondT} can be generalized to infinite order
\begin{align}
PT^\alpha|\alpha\rangle&=0,\quad \alpha\in\mathcal{M}_0, \label{CondTotal}
\end{align}
which states that the (connected) internal cluster operators have null actions on their own vacua.
This gives rise to the following GN
\begin{align}
\langle\beta|e^{T^\alpha}|\alpha\rangle&=\delta_{\beta\alpha}+ \sum_{n=2}\frac{1}{n!}\langle\beta|(T^\alpha)^n|\alpha\rangle,\quad \forall \alpha,\beta\in\mathcal{M}_0, \label{CondTotalMat}
\end{align}
which is clearly different from the IN \eqref{INC}, for products of external cluster operators can become internal.
Nevertheless, the intermediate and general normalizations are \emph{indistinguishable} up to the first order.

Given that the cluster operators are either internal or external for a closed model space $\mathcal{M}_X$
(termed ``special classes of incomplete model spaces'' in Ref. \citenum{Bartlett1989Mx}; vide post), the
T-condition \eqref{CondTotal}/\eqref{CondT} dictates that the amplitudes of all internal second-quantized operators in $T^\alpha$
are identically zero, so that the T-condition can in this case be extended to
\begin{align}
PT^\alpha|\beta\rangle&=0,\quad PT^{\alpha(i)} |\beta\rangle=0, \quad \forall \alpha,\beta\in\mathcal{M}_X, \label{CondTotalMx}
\end{align}
just like the case of CAS.

Under the above T-condition \eqref{CondTotal} or \eqref{CondTotalMx}, the cluster operators take the same form as
in single-reference CCSD
\begin{align}
T^{\alpha}_1&=\tilde{t}^I_A(\alpha)\{a^A_I\}_\alpha\nonumber\\
&=t_a^i(\alpha) \{a^a_i\}_\alpha + t_u^i(\alpha) \{a^u_i\}_\alpha + t_a^v(\alpha)\{a^a_v\}_\alpha+ \tilde{t}_{u}^v \{a^u_v\}_\alpha,\label{T1-Clusters}\\
T^{\alpha}_2&=\frac{1}{4}\tilde{t}_{AB}^{IJ}(\alpha)\{a^{AB}_{IJ}\}_\alpha\nonumber\\
&=\frac{1}{4} t_{ab}^{ij}(\alpha)\{a^{ab}_{ij}\}_\alpha + \frac{1}{2} t_{au}^{ij}(\alpha) \{a^{au}_{ij}\}_\alpha
+\frac{1}{4} t_{u_1u_2}^{ij}(\alpha)\{a^{u_1u_2}_{ij}\}_\alpha \nonumber\\
&+\frac{1}{2}t_{ab}^{iv}\{a^{ab}_{iv}\}_\alpha +\frac{1}{4}t_{ab}^{v_1v_2}\{a^{ab}_{v_1v_2}\}_\alpha+t_{au}^{iv}(\alpha)\{a^{au}_{iv}\}_\alpha\nonumber\\
&+\frac{1}{2} t_{au}^{v_1v_2}(\alpha)\{a^{au}_{v_1v_2}\}_\alpha + \frac{1}{2} t_{u_1u_2}^{iv}(\alpha)\{a^{u_1u_2}_{iv}\}_\alpha
+\frac{1}{4}\tilde{t}_{u_1u_2}^{v_1v_2}\{a_{v_1v_2}^{u_1u_2}\}_\alpha,\label{T2-Clusters}
\end{align}
where the tilde over $t$ emphasizes the exclusion of internal excitations.

\subsection{First-order amplitude equation} \label{SecFirstTeq}
The previous analysis shows that the diagonal zeroth-order Hamiltonian \eqref{Cond00OP} along with
the first- and second-order T-conditions \eqref{CondT} have to be invoked to obtain the desired first-order (Eq. \eqref{Veff1}) and second-order (Eq. \eqref{Veff2a}/\eqref{Veff2})
effective Hamiltonians. However, this zeroth-order Hamiltonian is unacceptable for the determination of the first-order amplitudes
$t_{l\alpha}^{(1)}$ associated with the external functions $|\chi_{l\alpha}\rangle$. To see this, we derive the first-order amplitude equation
by projecting $\langle\chi_{l\alpha}|$ on the left of Eq. \eqref{GenBlochT}:
\begin{align}
\langle\chi_{l\alpha}|Q\{V^\alpha + [H_0^\alpha, T^{\alpha(1)}]\}|\alpha\rangle&=
\sum_{\beta\in\mathcal{M}_0, \beta\ne\alpha} \langle\chi_{l\alpha}|Q\{T^{\beta(1)}-T^{\alpha(1)}\}|\beta\rangle \langle\beta|H_0^\alpha|\alpha\rangle,\label{FirstTeq}
\end{align}
which has a similar structure to Eq. \eqref{Heff1a}.
It is clear that the use of $H_0^\alpha$ \eqref{Cond00OP}
makes the right-hand side of Eq. \eqref{FirstTeq} vanish, so as to miss all $P$-space couplings. It follows that
different partitionings of the full Hamiltonian must be adopted for the cluster operators and the effective Hamiltonian.
As scrutinized previously\cite{PASPT2}, for a non-closed PAS $\mathcal{M}_0$, the connectivity of $t_{l\alpha}^{(1)}$
can only be ensured by using the following zeroth-order Hamiltonian
\begin{align}
H_0^\alpha&=H_{0,\mathrm{I}}^{\alpha}+H_{0,\mathrm{A}}^{\alpha},\quad \mathrm{A = H, X, M},\label{H0final}
\end{align}
where the orbital-class-diagonal and Hermitian inactive term $H_{0,\mathrm{I}}^{\alpha}$ reads
\begin{align}
H_{0,\mathrm{I}}^{\alpha}&= f_I^J(\alpha)a^I_J+ f_A^B(\alpha)a^A_B \nonumber\\
&=f_I^In_I(\alpha)+ f_I^J(\alpha)\{a^I_J\}_\alpha+ f_A^B(\alpha)\{a^A_B\}_\alpha,\label{H0Inactive}\\
f_p^q(\alpha)&=h_p^q+\bar{g}_{pI}^{qI}n_I(\alpha),\quad \bar{g}_{pq}^{rs}=g_{pq}^{rs}-g_{pq}^{sr}, \label{Fspin}
\end{align}
whereas the orbital-class-off-diagonal and non-Hermitian active term $H_{0,\mathrm{H}}^{\alpha}$ takes the following form
\begin{align}
H_{0,\mathrm{H}}^{\alpha}
&=\sum_{v,u}^{\prime} f_u^v(\alpha)\{a^u_v\}_{\alpha}
 +\frac{1}{2}\sum_{u_1,u_2,v_1,v_2}g_{u_1 u_2}^{v_1v_2}\{a^{u_1u_2}_{v_1v_2}\}_{\alpha}.\label{H0Active}
\end{align}
Here, the prime in the first summation is to exclude the particular case of
$|\beta\rangle=\{a_v^u\}_\alpha |\alpha\rangle$ \emph{and} $|\chi_{l\alpha}\rangle=\{a_{iv}^{au}\}_\alpha |\alpha\rangle$,
which would otherwise result in disconnected amplitudes $t^{\alpha(1)}_{ai,uv}$.
This particular choice of $H_{0,\mathrm{H}}^{\alpha}$ amounts to shifting the contribution of the connectivity-violating pairs (CVP)
$\{|\beta\rangle=\{a_v^u\}_\alpha |\alpha\rangle\in \mathcal{M}_0, \{a_{iv}^{au}\}_\alpha|\alpha\rangle\in\mathcal{M}_0^\perp \mbox{  s.t.  }
\{a_i^a\}_\alpha|\alpha\rangle\in\mathcal{M}_0\}$ to the second and higher orders,
which are always present in IN-GMS-SU-CCSD\cite{GMS-SU-CC2003}, so as to render the latter size-inextensive (as anticipated
by Nooijen and coworkers\cite{nooijen2005reflections}).
It can readily be checked that $QH_0^\alpha P_\alpha=0$, which is also a necessary condition for $H_0^\alpha$.
On the other hand, the fact that $[H_{0,\mathrm{H}}^{\alpha}, T^\alpha]=0$ merely simplifies the calculation
of the left-hand side of Eq. \eqref{FirstTeq}. It will be shown elsewhere\cite{PASPT2Eigen} that $H_{0,\mathrm{H}}^{\alpha}$
can be extended to
\begin{align}
H_{0,\mathrm{X}}^{\alpha}
&=\left[\sum_{v,u} f_u^v(\alpha)\{a^u_v\}_{\alpha}
 +\frac{1}{2}\sum_{u_1,u_2,v_1,v_2}g_{u_1 u_2}^{v_1v_2}\{a^{u_1u_2}_{v_1v_2}\}_{\alpha}\right]+\mathrm{h.c.}, \label{H0ActiveMx}
\end{align}
when $\mathcal{M}_0$ is itself a closed model space $\mathcal{M}_X$. In this case, both $\{a^u_v\}_{\alpha}$ and $\{a^{u_1u_2}_{v_1v_2}\}_{\alpha}$
are closed operators. Moreover, $[H_{0,\mathrm{X}}^{\alpha}, T^\alpha]$
no longer vanishes but provides couplings between the single and double amplitudes,
which are not present for $H_{0,\mathrm{H}}^{\alpha}$ in Eq. \eqref{H0Active}.

\begin{figure}[H]
	\centering
	\includegraphics[width=0.6\textwidth]{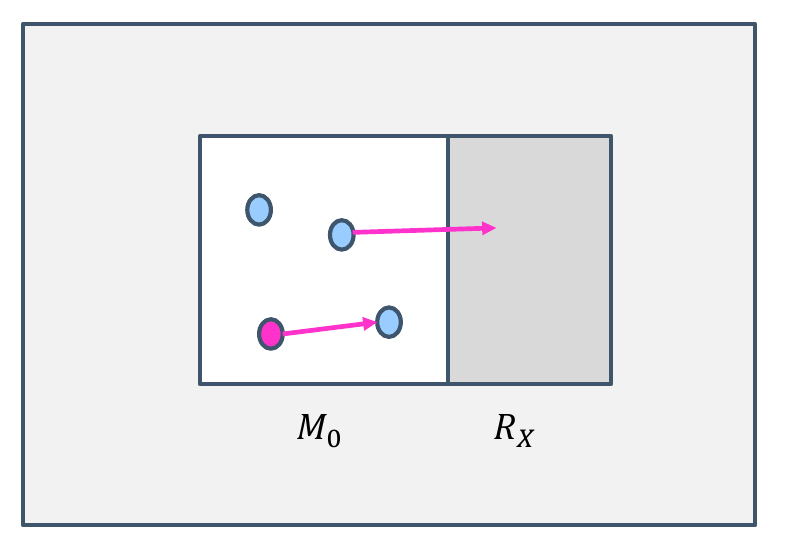}
	\caption{Illustration of closed, quasi-open and open operators for non-closed model space $\mathcal{M}_0$.
$\mathcal{M}_0\oplus\mathcal{R}_X=\mathcal{M}_X\subseteq\mathcal{M}_C, P=\mathcal{M}_0, Q=1-P=\mathcal{R}_X\oplus \mathcal{M}_X^\perp$.
Closed operators always stay in $\mathcal{M}_0$, open operators always go from $\mathcal{M}_0$ to $\mathcal{M}_X^\perp$,
whereas quasi-open operators can stay in $\mathcal{M}_0$ but can also go from $\mathcal{M}_0$ to $\mathcal{R}_X$.
 }
	\label{qOpen}
\end{figure}

To define a closed model space $\mathcal{M}_X$ ($=\mathcal{M}_0\oplus\mathcal{R}_X$), we borrow the concept of quasi-open operators\cite{MukerjeeClosedOp}.
A quasi-open operator is a normal-ordered second-quantized LASO-only operator $\{A_{\beta\alpha}\}_\alpha$ that
connects the Fermi vacuum $|\alpha\rangle$ with another reference DET $|\beta\rangle$ (i.e.,
$|\beta\rangle=\{A_{\beta\alpha}\}_\alpha|\alpha\rangle\in\mathcal{M}_0$), but can also excite at least one reference DET $|\gamma\rangle$ to $\mathcal{R}_X$
(mathematically, $\{A_{\beta\alpha}\}_\alpha|\gamma\rangle\ne|\rangle$ and $\{A_{\beta\alpha}\}_\alpha|\gamma\rangle \in\mathcal{R}_X$), see Fig. \ref{qOpen}.
For comparison, a closed operator only induces transitions within $\mathcal{M}_0$, while
an open operator only generates functions belonging to $\mathcal{M}_X^\perp$ by acting on $\mathcal{M}_0$.
When normal-ordered with respect to the same vacuum, the operators have the following properties:
product of two disjoint open/quasi-open operators can be open/quasi-open or closed (i.e., OO = O or C; QQ = Q or C),
product of two disjoint closed operators is closed (i.e., CC = C), product of a closed operator and another disjoint open operator is open (i.e., CO = O),
while product of a closed operator and another disjoint quasi-open operator can be quasi-open or closed (i.e., CQ = QC = Q or C). 
Following these rules, the closed counterpart $\mathcal{M}_X$ of $\mathcal{M}_0$ can readily be generated
by a minimal extension of $\mathcal{M}_0$ until that all quasi-open operators just disappear.
In general, $\mathcal{M}_X$ is a subspace of $\mathcal{M}_C$ (i.e., $\mathcal{M}_0\subset\mathcal{M}_X\subseteq\mathcal{M}_C=\mathcal{M}_X+\mathcal{R}_C$).
It deserves to be emphasized that the closure of $\mathcal{M}_X$ is independent of the choice of vacuum
but is a property of $\mathcal{M}_X$ itself. Most importantly, second-quantized operators are either closed
($\hat{I}_I$) or open ($\hat{X}_I$) for $\mathcal{M}_X$
(i.e., $\{\hat{I}_I\}\mathcal{M}_X=\mathcal{M}_X$, $\{\hat{X}_I\}\mathcal{M}_X=\mathcal{M}_X^\perp$, $\hat{I}_I\hat{X}_J=\hat{X}_J\hat{I}_I$,
$\hat{I}_I\hat{I}_J=\hat{I}_K$, $\hat{X}_I\hat{X}_J=\hat{X}_K$)\cite{Bartlett1989Mx},
which is the very basis of the CAS-like T-condition \eqref{CondTotalMx}.
Because of this, it can be shown\cite{PASPT2Eigen} that $\mathcal{M}_X$ possesses the same property as $\mathcal{M}_C$
with respect to the connectedness of the energies obtained by diagonalization of the effective Hamiltonian \eqref{HeffFinal}.

It was attempted\cite{MukerjeeClosedOp,Mukerjee1989M0,Bartlett1990M0} long ago to formulate a size-extensive MRCC by
staying within the non-closed $\mathcal{M}_0$ characterized by the projector $P$. This proceeds by
requiring the effective Hamiltonian $H^{\mathrm{eff}}$ to be a closed operator on $\mathcal{M}_0$, meaning that
all matrix elements $H^{\mathrm{eff}}_{\beta\alpha}$ are enforced to vanish when
$\{|\beta\rangle\}$ (denoted collectively as qRef) are connected with
the vacuum $|\alpha\rangle$ by quasi-open operators $\{\{A_{\beta\alpha}\}_\alpha\}$.
This requirement assigns amplitudes to the qRefs which are otherwise vanishing due to the T-condition \eqref{CondTotal}.
Specifically, the qRefs associated with $|\alpha\rangle$ are to be promoted to the extended external space $\tilde{Q}_\alpha$
of $|\alpha\rangle$, viz.\cite{Mukerjee1989M0,Bartlett1990M0}
\begin{align}
\tilde{Q}_\alpha=Q+\bar{P}_\alpha,\quad  \bar{P}_\alpha=\sum_{\beta\in\mathcal{M}_0} |\beta\rangle\langle\beta|_{Q\{A_{\beta\alpha}\}_\alpha P\ne 0},\label{qQop}
\end{align}
which gives the corresponding reduced $P$-space projector and T-condition
\begin{align}
&\tilde{P}_\alpha=1-\tilde{Q}_\alpha=P+Q-\tilde{Q}_\alpha=P-\bar{P}_\alpha,\\
&\tilde{P}_\alpha T^{\alpha(i)}|\alpha\rangle=0 \Rightarrow T^\alpha|\alpha\rangle=\tilde{Q}_\alpha T^\alpha|\alpha\rangle. \label{qT-cond}
\end{align}
The amplitude equation \eqref{FirstTeq} is then revised to
\begin{align}
\langle\chi_{l\alpha}|\tilde{Q}_\alpha\{V^\alpha + [H_0^\alpha, T^{\alpha(1)}]\}|\alpha\rangle&=
\sum_{\beta\in\mathcal{M}_0, \beta\ne\alpha} \langle\chi_{l\alpha}|\tilde{Q}_\alpha\{T^{\beta(1)}-T^{\alpha(1)}\}|\beta\rangle \langle\beta|\tilde{P}_\alpha H_0^\alpha|\alpha\rangle,\label{qFirstTeq}
\end{align}
where $H_0^\alpha$ is composed of the inactive term $H_{0,\mathrm{I}}^\alpha$ \eqref{H0Inactive} and the following active term
\begin{align}
H_{0,\mathrm{M}}^{\alpha}
&=\left[\sum_{v,u}^{(\mathrm{cl})} f_u^v(\alpha)\{a^u_v\}_{\alpha}
 +\frac{1}{2}\sum^{(\mathrm{cl})}_{u_1,u_2,v_1,v_2}g_{u_1 u_2}^{v_1v_2}\{a^{u_1u_2}_{v_1v_2}\}_{\alpha}\right]+\mathrm{h.c.}. \label{H0ActiveMM}
\end{align}
Here, the superscript `(cl)' in the summations is to emphasize that both $\{a^u_v\}_{\alpha}$ and $\{a^{u_1u_2}_{v_1v_2}\}_{\alpha}$ are closed operators
for $\mathcal{M}_0$ (i.e., both $\{a^u_v\}_{\alpha}|\alpha\rangle$ and $\{a^{u_1u_2}_{v_1v_2}\}_{\alpha}|\alpha\rangle$ must
belong to $\tilde{P}_\alpha$). This is dictated by the condition $\tilde{Q}_\alpha H_{0,\mathrm{M}}^\alpha |\alpha\rangle=0$
that must be fulfilled by $H_{0,\mathrm{M}}^{\alpha}$ itself. Moreover, this particular choice is based on
the following analysis of the only potentially CVP
$\{|\beta\rangle=\{a_v^u\}_\alpha |\alpha\rangle\in \mathcal{M}_0, \{a_{iv}^{au}\}_\alpha|\alpha\rangle\in\mathcal{M}_0^\perp \mbox{  s.t.  }
\{a_i^a\}_\alpha|\alpha\rangle\in\mathcal{M}_0\}$:
\begin{enumerate}[(A)]
\item If $|\beta\rangle$ does not belong to $\tilde{P}_\alpha$, the right-hand side of Eq. \eqref{qFirstTeq} is identically zero,
irrespective of the form of $H_{0,\mathrm{M}}^\alpha$.
In this case, $\{a_v^u\}_\alpha$ is a quasi-open operator, so that $|\alpha\rangle$ will acquire an uncoupled amplitude $t^{\alpha(1)}_{ai,uv}$
when $\{a_{iv}^{au}\}_\alpha$ is also quasi-open (i.e., when $\{a_{iv}^{au}\}_\alpha|\alpha\rangle$ belongs to $\bar{P}_\alpha$).

\item\label{C-O} If $|\beta\rangle$ belongs to $\tilde{P}_\alpha$ and $\{a_{iv}^{au}\}_\alpha|\alpha\rangle$ belongs to $Q=1-P$,
then $\{a_v^u\}_\alpha$ and $\{a_{iv}^{au}\}_\alpha\equiv\{a_i^a\}_\alpha\{a_v^u\}_\alpha$ are closed and open operators for $\mathcal{M}_0$, respectively.
It follows that  $\{a_{i}^{a}\}_\alpha$ must be an open operator. Since the indices
$i$ (occupied) and $a$ (unoccupied) are by definition common to $|\alpha\rangle$ and $|\beta\rangle$,
$\{a_{i}^{a}\}_\alpha$ is identical to $\{a_{i}^{a}\}_\beta$. Therefore, the right-hand side of Eq. \eqref{qFirstTeq}
would be equal to $[t^{\beta(1)}_{a,i} - t^{\alpha(1)}_{a,i}]f_u^v(\alpha)$,
arising from the first term of $H_{0,\mathrm{M}}^\alpha$ \eqref{H0ActiveMM}. This is a connected quantity\cite{Bartlett1990M0}, for the common terms
in $t^{\beta(1)}_{a,i}$ and $t^{\alpha(1)}_{a,i}$ are cancelled out,
thereby left over only terms that are labeled with indices $u$ and $v$.
This is a fundamental difference between Eqs. \eqref{qFirstTeq} and \eqref{FirstTeq}: the latter
does not require that $\{a_{i}^{a}\}_\alpha$ be open, so that $\{a_{i}^{a}\}_\alpha|\alpha\rangle$
may belong to $P$ with zero $t^{\alpha(1)}_{a,i}$ [cf. Eq. \eqref{CondT}], thereby resulting in disconnected
$t^{\beta(1)}_{a,i}f_u^v(\alpha)$. As such, $H_{0,\mathrm{H}}^\alpha$ in Eq. \eqref{H0Active}
must be used in Eq. \eqref{FirstTeq} to enforce this to vanish.

\item\label{C-qO} If $|\beta\rangle$ belongs to $\tilde{P}_\alpha$ but $\{a_{iv}^{au}\}_\alpha|\alpha\rangle$ belongs to $\bar{P}_\alpha$,
then $\{a_v^u\}_\alpha$ and $\{a_{iv}^{au}\}_\alpha\equiv\{a_i^a\}_\alpha\{a_v^u\}_\alpha$ are closed and quasi-open operators for $\mathcal{M}_0$, respectively.
As for Case (\ref{C-O}), $\{a_i^a\}_\alpha$ is identical to $\{a_i^a\}_\beta$ and is quasi-open
(which implies that $\{a_i^a\}_\alpha|\alpha\rangle$ must be in $P$ and meanwhile belong to $\bar{P}_\alpha$). It is just that
$i$ and $a$ should be replaced with $v_1$ ($\ne v$) and $u_1$ ($\ne u$), respectively, to stick to our convention.
The simplest $\mathcal{M}_0$ for this case takes the following form 
\begin{align}
\mathcal{M}_0=\{ |\alpha\rangle, \{a_v^u\}_\alpha|\alpha\rangle=|\beta\rangle, \{a_{v_1}^{u_1}\}_\alpha|\alpha\rangle, \{a_{vv_1}^{uu_1}\}_\alpha|\alpha\rangle \} \oplus \{\{a_{v_2}^{u_2}\}_\alpha, \{a_{vv_2}^{u u_2}\}_\alpha\},
\end{align}
where the first part is a closed subspace, whereas the components in the second part ensure that
$\{a_v^u\}_\alpha$ is closed, while both $\{a_{v_1}^{u_1}\}_\alpha$ and $\{a_{vv_1}^{uu_1}\}_\alpha$ are quasi-open.
On the other hand, when $|\beta\rangle$ is taken as the vacuum,  $\{a_{v_1}^{u_1}\}_\beta$ [$=\{a_{v_1}^{u_1}\}_\alpha$] is quasi-open, so that   
$\{a_{v_1}^{u_1}\}_\beta |\beta\rangle$ [$=\{a_{vv_1}^{uu_1}\}_\alpha|\alpha\rangle$] belongs to $\bar{P}_\beta$.
Therefore, both $|\alpha\rangle$ and $|\beta\rangle$ will gain amplitudes thanks to their extended external spaces [cf. \eqref{qQop}].  
Without going into further details, the amplitudes determined by Eq. \eqref{qFirstTeq} are also connected in this case. 
\end{enumerate}

Having determined the connected first-order amplitudes, we can define the lowest-order effective Hamiltonians as
\begin{align}
H_{\beta\alpha}^{\mathrm{eff}(0)}&=\langle\beta|\tilde{P}_\alpha H_0^\alpha|\alpha\rangle,\label{qVeff0}\\
H_{\beta\alpha}^{\mathrm{eff}(1)}&=\langle\beta|\tilde{P}_\alpha V^\alpha|\alpha\rangle,\label{qVeff1}\\
H_{\beta\alpha}^{\mathrm{eff}(2)}&=\langle\beta|\tilde{P}_\alpha[V^\alpha, T^{\alpha(1)}]|\alpha\rangle
-\sum_{\gamma\in \mathcal{M}_0, \gamma\ne\alpha}\langle\beta|\tilde{P}_\alpha (T^{\gamma(1)}-T^{\alpha(1)})|\gamma\rangle H_{\gamma\alpha}^{\mathrm{eff}(1)}. \label{qVeff2a}
\end{align}
The renormalization term on the right-hand side of Eq. \eqref{qVeff2a} is connected, unlike that in Eq. \eqref{Veff2a} for a non-closed $\mathcal{M}_0$.
Nevertheless, it is difficult to compute (due to the crossed term $T^{\alpha(1)}|\gamma\rangle|_{\alpha\ne\gamma}$) and numerically insignificant\cite{PASPT2Eigen}.
Ignoring this term gives rise to  the following effective Hamiltonian $H^{\mathrm{eff}[2]}_M$
\begin{align}
H^{\mathrm{eff}[2]}_M&=P \left[ \sum_{\alpha\in\mathcal{M}_0}\tilde{P}_\alpha \{ H+[H, T^{\alpha(1)}] \} P_\alpha \right] P, \label{qHeff}
\end{align}
which complies with the imposed condition on the qRefs
\begin{align}
\bar{P}_\alpha H^{\mathrm{eff}[2]}_M|\alpha\rangle=0.\label{qHeffCond}
\end{align}
It is this particular structure of the effective Hamiltonian that ensures the connectedness of
the energies upon diagonalization\cite{Mukerjee1989M0,Bartlett1990M0}.


To illustrate the above, consider the following three-dimensional $\mathcal{M}_0$
\begin{align}
\mathcal{M}_0=\mathrm{span}\{|\alpha\rangle, |\beta\rangle=\{a_{v_1}^{u_1}\}_\alpha |\alpha\rangle,
|\gamma\rangle=\{a_{v_2}^{u_2}\}_\alpha |\alpha\rangle\}|_{v_1\ne v_2, u_1\ne u_2}, \label{M0}
\end{align}
which takes $|\alpha\rangle$ as the vacuum. It is readily seen that
both $A_{\beta\alpha}=\{a_{v_1}^{u_1}\}_\alpha$ and $A_{\gamma\alpha}=\{a_{v_2}^{u_2}\}_\alpha$ are quasi-open operators
that generate a nonvanishing DET $\{a_{v_1v_2}^{u_1u_2}\}_\alpha |\alpha\rangle$ outside $\mathcal{M}_0$, so that
the projectors associated with $|\alpha\rangle$ are to be reduced to
\begin{align}
\tilde{Q}_\alpha&=Q+\bar{P}_\alpha,\quad \bar{P}_\alpha=P_\beta+P_\gamma,\quad \tilde{P}_\alpha=P_\alpha.
\end{align}
In contrast, when $|\beta\rangle$ or $|\gamma\rangle$ is taken as the vacuum, the $\mathcal{M}_0$ \eqref{M0} has no quasi-open operators.
Therefore, the 3-by-3 Hamiltonian matrix $\mathbf{H}^{\mathrm{eff}[2]}_M$ has in this case the following structure
\begin{align}
\mathbf{H}^{\mathrm{eff}[2]}_M&=\begin{pmatrix} E_{11} & E_{12}&E_{13}\\
0 & E_{22}& E_{23}\\
0 & E_{32}& E_{33}\end{pmatrix},\label{HMstructure}
\end{align}
where $E_{11}$ is equal to $E^{[2]}_\alpha$ \eqref{MP2Ene}. It is seen that the vacuum $|\alpha\rangle$ is completely decoupled from
the rest components of $\mathcal{M}_0$ in the sense of static correlation. It can hence be deduced that
this Ansatz, even at the MRCCSD level, has only limited accuracy in general, although preliminary applications look promising\cite{BartlettLiHM0,MukherjeeNullBlock}.
What is more severe is that the existence of null blocks
makes spin adaptation of $H^{\mathrm{eff}[2]}_M$ \eqref{qHeff} generally impossible.


\subsection{Remarks}

\begin{figure}[H]
	\centering
	\includegraphics[width=0.8\textwidth]{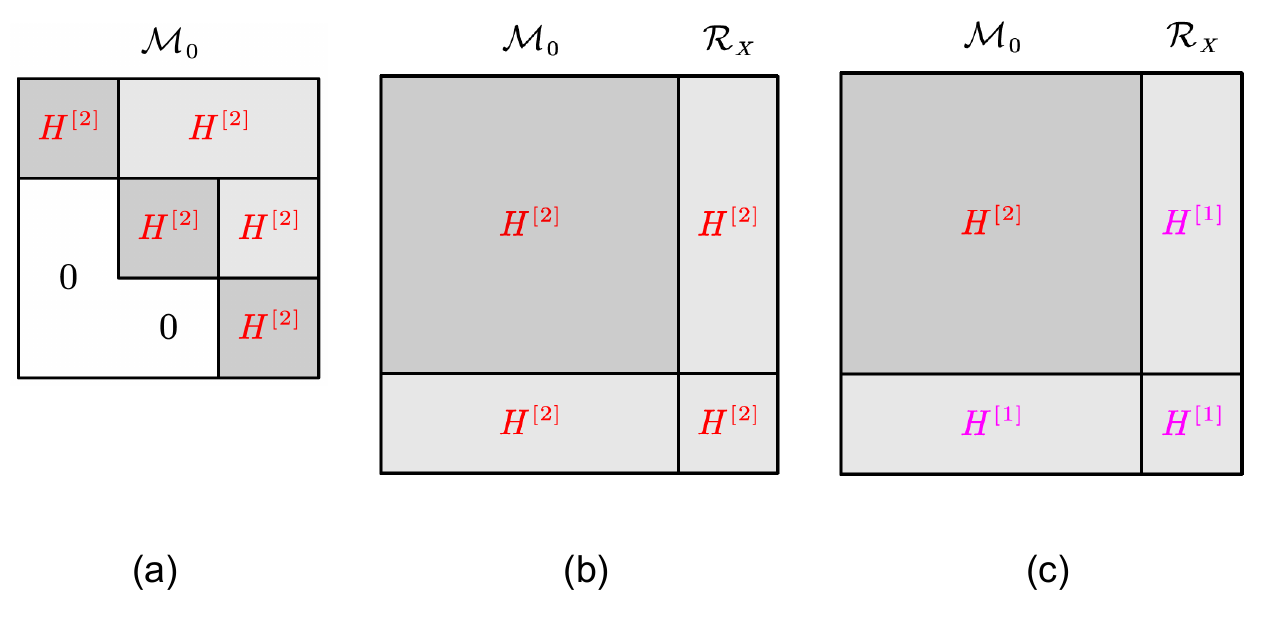}
	\caption{Structures of (a) PASPT2M, (b) PASPT2X and (c) PASPT2H Hamiltonians. $\mathcal{M}_0\oplus\mathcal{R}_X=\mathcal{M}_X$.
Note that the block upper triangular structure (a) is typical for PASPT2M, although not universal.
 }
	\label{Hstructure}
\end{figure}

\begin{figure}[H]
	\centering
	\includegraphics[width=0.8\textwidth]{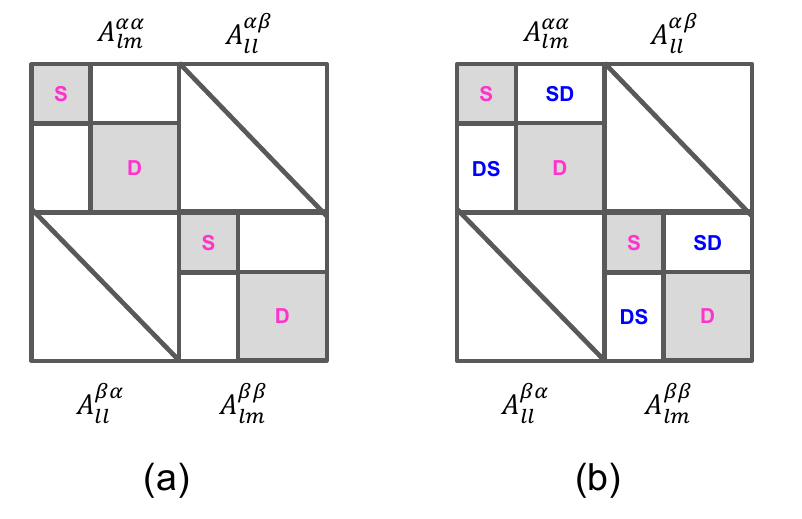}
	\caption{Structures of the $\mathbf{A}$ matrices in Eq. \eqref{LinearSys} for PASPT2H (a) and PASPT2X/PASPT2M (b).
Unfilled areas are null. }
	\label{Amat}
\end{figure}

Several remarks are in order.
\begin{enumerate}[(I)]
\item Three variants of PASPT2 have been discussed thus far, see Table \ref{PASPT2variant}.
In the variant defined by Eq. \eqref{qFirstTeq} along with
Eqs. \eqref{H0Inactive} and \eqref{H0ActiveMM} for the amplitudes as well as Eq. \eqref{qHeff} for the effective
Hamiltonian $H^{\mathrm{eff[2]}}_M$, both the cluster operators and effective Hamiltonian are defined on a non-closed PAS $\mathcal{M}_0$.
This variant can hence be coined as PASPT2M.
In contrast, in the variant defined by Eq. \eqref{FirstTeq}
along with Eqs. \eqref{H0Inactive} and \eqref{H0ActiveMx} for the amplitudes as well as Eq. \eqref{HeffFinal} for the effective
Hamiltonian $H^{\mathrm{eff[2]}}_X$, both the cluster operators and effective Hamiltonian are defined on $\mathcal{M}_X$, the closed counterpart of $\mathcal{M}_0$.
This variant hence deserves the name of PASPT2X. Both PASPT2M and PASPT2X are strictly size-extensive (see Sec. \ref{SecGSE}). However,
both are beset with the intruder-state problem.
Our original variant\cite{PASPT2} is defined by Eq. \eqref{FirstTeq} along with
Eqs. \eqref{H0Inactive} and \eqref{H0Active} for the amplitudes of the cluster operators defined on $\mathcal{M}_0$, as well as
the following \emph{intermediate} Hamiltonian defined on $\mathcal{M}_X$
\begin{align}
(H^{\mathrm{eff}[2]}_H)_{\beta\alpha}&=
\begin{cases} H^{\mathrm{eff}[2]}_{\beta\alpha},\quad \beta \mbox{ and } \alpha\in\mathcal{M}_0,\\
H_{\beta\alpha},\quad \beta  \mbox{ or } \alpha\in \mathcal{R}_X=\mathcal{M}_X\ominus\mathcal{M}_0.
\end{cases}\label{HeffH}
\end{align}
That is, the $\mathcal{R}_X$ space complementary to $\mathcal{M}_0$ within $\mathcal{M}_X$ is taken here as a buffer, which is effected by
the extended T-condition\cite{PASPT2}
\begin{subequations}\label{extended-T}
\begin{equation}
\langle \beta | T^{\alpha(1)} |\alpha\rangle = 0, \quad \forall \beta\in \mathcal{M}_X, \quad \alpha\in \mathcal{M}_0,
\end{equation}
\begin{equation}
T^{\beta(1)}|\beta\rangle=0,\quad \forall \beta\in \mathcal{R}_X=\mathcal{M}_X\ominus\mathcal{M}_0.
\end{equation}
\end{subequations}
Literally, only those functions belonging to $\mathcal{M}_0$ are allowed to be excited to $\mathcal{M}_X^\perp$.
This has a sound physical basis: by construction, the functions belonging to $\mathcal{R}_X$ do not contribute significantly to the desired states
but are prone to intruder states, so that they should not be perturbed.
Conceptually, this variant looks like a hybrid of PASPT2M and PASPT2X
and can hence be termed PASPT2H. Compared with PASPT2X, PASPT2H merely neglects the second-order corrections
[cf. the second term of Eq. \eqref{HeffFinal} and  Fig. \ref{Hstructure}] that are
important only for states having major projections on $\mathcal{R}_X$, given the minor differences
between the PASPT2H and PASPT2X amplitudes (which, as already said, is guaranteed by a well-chosen $\mathcal{M}_0$). On the other hand,
owing to its extended Hamiltonian and reduced non-Hermiticity, PASPT2H is intrinsically more accurate than PASPT2M.
Nevertheless, the $H^{\mathrm{eff}[2]}_H$ Hamiltonian \eqref{HeffH} does not satisfy all conditions for the diagonal and off-diagonal
Hamiltonian matrix elements that are essential for $H^{\mathrm{eff}[2]}_X$ to deliver $|M_X|$ connected energies\cite{PASPT2Eigen}.
As such, PASPT2H is not strictly size-extensive per se, unlike PASPT2X. Nevertheless, the PASPT2H energies for the $|M_0|$ states
with major projections on $\mathcal{M}_0$ are near-extensive, as confirmed numerically\cite{PASPT2} (see also Sec. \ref{SecGSE}). Anyway,
the approximate treatment (to first order) of those states with major projections on $\mathcal{R}_X$ is physically sound,
for a finite one-particle basis usually does not support such high-lying states.
\item The amplitude equation \eqref{qFirstTeq} for PASPT2M can be recast into a system of linear equations
\begin{align}
\mathbf{A}\mathbf{t}=\mathbf{V},\label{LinearSys}
\end{align}
where
\begin{align}
A_{l\alpha,m\alpha}&=-\Delta E^{(0)}_{l\alpha}\delta_{lm} - (1-\delta_{lm}) H^{(0)}_{l\alpha,m\alpha},\label{A_matrix_elmt_1}\\
 A_{l\alpha,l\beta}&=\langle\beta|\tilde{P}_\alpha H_0^{\alpha}|\alpha\rangle, \quad \beta\ne\alpha\in\mathcal{M}_0,\label{A_matrix_elmt_2}\\
\Delta E_{l\alpha}^{(0)}&=H^{(0)}_{l\alpha,l\alpha}-\langle\alpha|H^{\alpha}_0|\alpha\rangle,\label{denorminator}\\
 H^{(0)}_{l\alpha,m\alpha}&=\langle\chi_{l \alpha}|\tilde{Q}_\alpha H_0^\alpha \tilde{Q}_\alpha|\chi_{m \alpha}\rangle,\label{QH0Qmat}\\
 V_{l\alpha}&=\langle\chi_{l\alpha}|\tilde{Q}_\alpha V^\alpha|\alpha\rangle.\label{Vmat}
\end{align}
The counterpart \eqref{LinearSys} of the amplitude equation \eqref{FirstTeq} for PASPT2X and PASPT2H is obtained
by replacing $\tilde{P}_\alpha$ and $\tilde{Q}_\alpha$ in Eqs. \eqref{A_matrix_elmt_2}, \eqref{QH0Qmat}
and \eqref{Vmat} with $P$ and $Q$, respectively.
Moreover, $\mathcal{M}_0$ in Eq. \eqref{A_matrix_elmt_2} should be replaced with $\mathcal{M}_X$ in the case of PASPT2X,
while $H_0^\alpha$ in Eq. \eqref{QH0Qmat} should be replaced with $H_{0,\mathrm{I}}^\alpha$ in the case of PASPT2H (see discussions
below equation (69) in Ref. \cite{PASPT2arXiv}).
Both $\{t_{l\alpha}\}$ and $\{V_{l\alpha}\}$ are understood as column vectors, whereas $\mathbf{A}$ is a very sparse matrix (see Fig. \ref{Amat}):
off-diagonal in the $Q$-space but diagonal in the $P$-space [cf. Eq. \eqref{A_matrix_elmt_1}] or off-diagonal in the $P$-space
but diagonal in the $Q$-space [cf. Eq. \eqref{A_matrix_elmt_2}]. In other words, there exist explicit couplings only among
different $Q$-space functions originated from the same $P$-space function or among different $P$-space functions yielding the same $Q$-space function.
There exists a subtle difference between the $\mathbf{A}$ matrix for PASPT2H and that for PASPT2X and PASPT2M:
$H_{0,\mathrm{H}}^\alpha$ in Eq. \eqref{H0Active} does not couple single and double excitations from the same $P$-space function, whereas
$H_{0,\mathrm{X}}^\alpha$/$H_{0,\mathrm{M}}^\alpha$ in Eq. \eqref{H0ActiveMx}/\eqref{H0ActiveMM} (more precisely, the one-body term therein) does so.
It is obvious that the number of unknown amplitudes $\{t_{l\alpha}\}$ is equal to that of the determining conditions
$(\mathbf{At})_{l\alpha}=V_{l\alpha}$, so that Eq. \eqref{LinearSys} has a unique solution, provided that $\mathbf{A}$ is nonsingular
(which is always the case for PASPT2H).

\item The computation is truly simple for all three variants:
the MP1 amplitudes $t^0_{l\alpha}=-V_{l\alpha}/\Delta E^{(0)}_{l\alpha}$ for each reference $|\alpha\rangle$
can be computed in an embarrassingly parallel fashion.
The coupled amplitudes $\mathbf{t}$ can simply be expanded as a
linear combination of the uncoupled $\mathbf{t}^0$-amplitudes (LUCT, i.e.,
$\mathbf{t}=\sum_\alpha\mathbf{t}^0_\alpha \mathbf{c}_\alpha$), with the expansion coefficients determined
by minimizing the cost function $\|\mathbf{A}\mathbf{t}- \mathbf{V}\|_2$, which is just a non-iterative least-squares fitting.
This LCUT approximation is sufficiently accurate for most purposes. Otherwise, Krylov subspace iterations
can further be invoked to obtain fully convergent amplitudes. In total $|\mathcal{M}_0|$ or $|\mathcal{M}_X|$ states
can be obtained by diagonalizing the non-Hermitian Hamiltonian matrices
$\mathbf{H}^{\mathrm{eff}[2]}_H$/$\mathbf{H}^{\mathrm{eff}[2]}_M$ or $\mathbf{H}^{\mathrm{eff}[2]}_X$.
\item Additional merits of PASPT2 are as follows:
(1) spin-adaptation can readily be achieved; (2) its combination with relativistic-QED Hamiltonians\cite{eQED,LiuPerspective2020} or
spin-separated X2C Hamiltonians\cite{X2CSOC1,X2CSOC2} is straightforward, following our previous works\cite{4C-iCIPT2,SOiCI,MetaWave};
(3) its state-universality renders it an ideal model for machine learning of dressed active-space Hamiltonians; (4) its near-extensivity
enables its application to strongly correlated solid states.

\end{enumerate}

\begin{table}
	\centering
	\caption{Comparison of variants of PASPT2 }
\tiny
	\begin{threeparttable}
		\begin{tabular}{clcccrcc}\toprule
	variant	& &$H_0^\alpha$                              & $T^{\alpha(1)}$ &$H^{\mathrm{eff[2]}}$ &\multicolumn{1}{c}{size-extensive} & size-consistent&intruder-free\\\toprule
PASPT2M&equation&\eqref{H0final}, \eqref{H0Inactive}, \eqref{H0ActiveMM}&\eqref{qFirstTeq}
                                                         &\eqref{qHeff}         &rigorous          &rigorous& no\\
       &domain\tnote{a}  &  full                         &$\mathcal{M}_0$  &$\mathcal{M}_0$       &                  & \\
       &property         & additively separable          & connected       &connected \& closed   &                  &  \\
PASPT2X&equation&\eqref{H0final}, \eqref{H0Inactive}, \eqref{H0ActiveMx}&\eqref{FirstTeq} &\eqref{HeffFinal}
                                                         &rigorous          &rigorous& no\\
       &domain\tnote{a}  &  full                         &$\mathcal{M}_X$  &$\mathcal{M}_X$       &                  & \\
       &property         & additively separable          & connected       &connected \& closed   &                  &  \\
PASPT2H&equation&\eqref{H0final}, \eqref{H0Inactive}, \eqref{H0Active}  &\eqref{FirstTeq} &\eqref{HeffH}
                                                         &near-rigorous     &rigorous& yes\\
       &domain\tnote{a}  &  full                         &$\mathcal{M}_0$  &$\mathcal{M}_X$       &                  & \\
       &property         & additively separable          & connected       &connected \& closed   &                  &  \\
       \bottomrule
		\end{tabular}
	\begin{tablenotes}
            \item[a]$\mathcal{M}_0\oplus\mathcal{R}_X=\mathcal{M}_X$.
		\end{tablenotes}
\end{threeparttable} \label{PASPT2variant}
\end{table}

\section{Test of generalized size-extensivity}\label{SecGSE}
Unlike size-consistency, size-extensivity cannot be tested numerically for systems with interacting but non-repeating units. To resolve this,
Nooijen and coworkers\cite{nooijen2005reflections} introduced an operational definition of size-extensivity, generalized size-extensivity (GSE).
It proceeds by first defining a second-quantized Hamiltonian in normal order with respect to a reference $|\alpha\rangle$
\begin{align}
H&=H_0^\alpha + V^\alpha, \\
H_0^\alpha&=\langle\alpha|H|\alpha\rangle+f_p^q(\alpha)\{a^p_q\}_\alpha,\\
 V^\alpha&=\frac{1}{2}g_{pq}^{rs}\{a^{pq}_{rs}\}_\alpha,\label{V_RS}
\end{align}
and then partitioning the whole set of orbitals into two disjoint subsets $R$ and $S$ in an arbitrary manner,
so as to split the two-body operator $V^\alpha$ \eqref{V_RS} into three terms
\begin{align}
V^\alpha&=V^\alpha_R+V^\alpha_S+V^\alpha_{RS},\\
V^\alpha_R&=\frac{1}{2}g_{pq}^{rs}\{a^{pq}_{rs}\}_\alpha,\quad p,q,r,s\in R,\\
V^\alpha_S&=\frac{1}{2}g_{pq}^{rs}\{a^{pq}_{rs}\}_\alpha,\quad p,q,r,s\in S,\\
V^\alpha_{RS}&=V^\alpha-V^\alpha_R-V^\alpha_S.\label{Vmix}
\end{align}
After switching off the crossed operator $V^\alpha_{RS}$ \eqref{Vmix} (by zeroing all two-electron integrals $g_{pq}^{rs}$ with mixed orbitals),
the following calculations and state-assignment are performed for a given method:
\begin{enumerate}[(1)]
\item Perform calculation with $V^\alpha_R$ and $V^\alpha_S$, giving energies $E_{R/S}$.
\item Perform calculation with $V^\alpha_R$ and $V^\alpha_S=0$, giving energies $E_{R/S_0}$.
\item Perform calculation with $V^\alpha_S$ and $V^\alpha_R=0$, giving energies $E_{R_0/S}$.
\item Perform calculation with $V^\alpha_R=0$ and $V^\alpha_S=0$, giving energies $E^0_{R/S}$.
\item Align the states of calculations (1) to (3) to those of (4) by compositions, and calculate the extensivity error
$\Delta^{\mathrm{GSE}}=E_{RS}-(E_{R/S_0}+E_{R_0/S}-E^0_{R/S})$ for each state.
\end{enumerate}
It is of interest to see that the two-electron integrals in the fluctuation potential $V^\alpha$ \eqref{V_RS}
are reference-independent, so that methods (such as PASPT2) working with multiple partitionings of the Hamiltonian
can also be put to test. However, the GSE procedure does not refer to the nature of model spaces.
That is, the orbital partitioning may always be possible for an arbitrary PAS $\mathcal{M}_0$.
Moreover, it is impractical to consider all possible orbital partitionings.
Therefore, in the context of PASPT2, we can only consider some special cases.
It appears that the model spaces $\mathcal{M}_0$ considered by Li and Paldus\cite{GMS-SU-CCSDJCP2004}
for equilibrium water with the cc-pVDZ basis set\cite{Dunning1989} can serve as a good showcase for the present purpose.
As can be seen from Table \ref{table:ModelSpaceInfo},
they consist of 10, 12, 8 and 14 determinants for the $A_1$, $A_2$, $B_1$ and $B_2$ irreducible representations (irrep) of $C_{2v}$, respectively,
and support in total 22 singlet states.
It follows that the canonical Hartree-Fock (HF) orbitals belonging to the $a_1$ and $a_2$ irreps
can be assigned to subset $R$, while those belonging to the $b_1$ and $b_2$ irreps to subset $S$.
This is also the orbital partitioning considered by Nooijen and coworkers\cite{nooijen2005reflections}.
The corresponding minimal closed spaces $\mathcal{M}_X$ consist\cite{PASPT2} of 104, 22, 20 and 104 determinants for
the $A_1$, $A_2$, $B_1$ and $B_2$ irreps, respectively,
which are much smaller than the corresponding sizes of CAS(8,8).

\begin{figure}[H]
	\centering
	\includegraphics[width=0.6\textwidth]{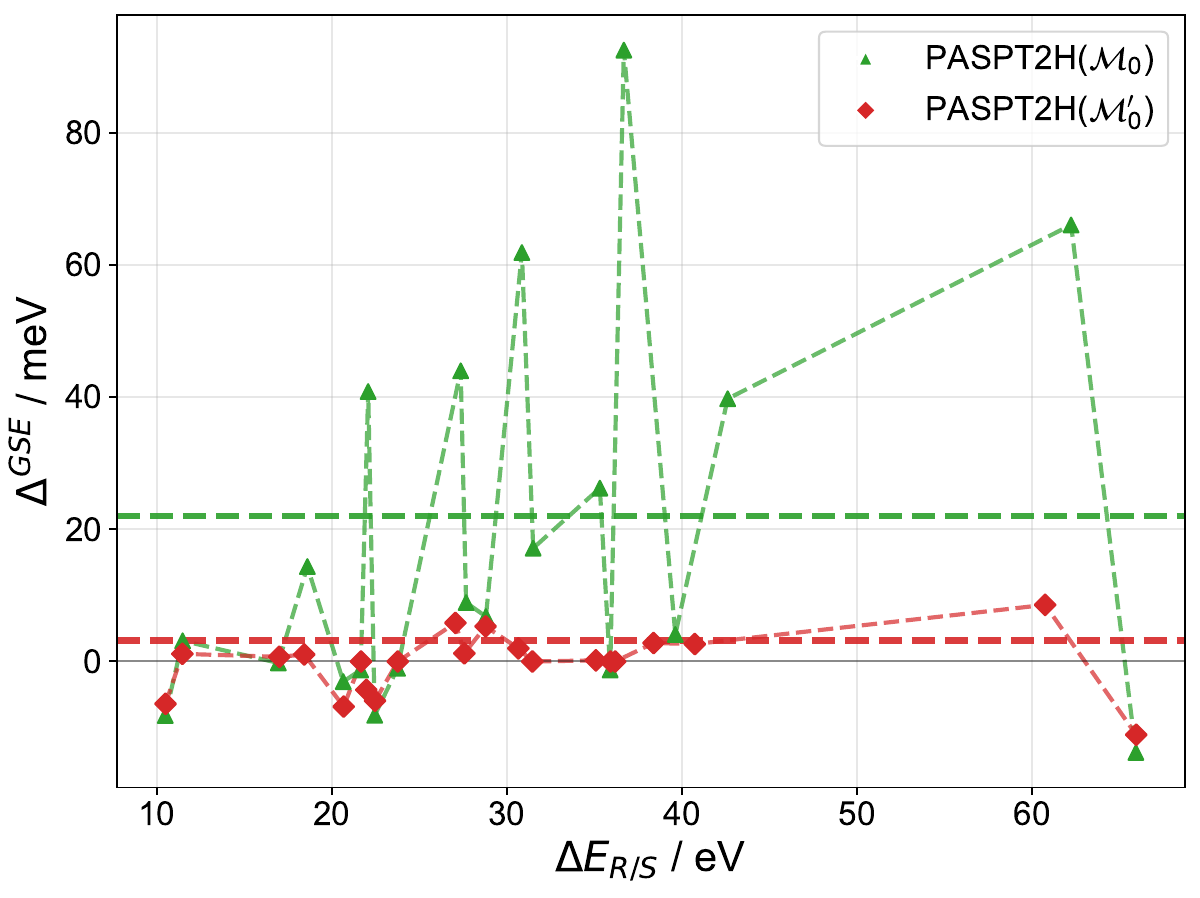}
	\caption{Size-extensivity errors of PASPT2H with different model spaces (see Table \ref{table:ModelSpaceInfo}).
 }
	\label{GSE}
\end{figure}

Given the above background, we first check the GSE of PASPT2M. As can be seen from Table \ref{tab:paspt2m-results},
PASPT2M is indeed strictly size-extensive.
In contrast, PASPT2H is not size-extensive, see Table \ref{tab:paspt2h-results}. However, the size-extensivity errors
are merely 1$\sim$ 40 meV for the lowest 13 states (ordered according to $E_{R/S}$). Given that $|\mathcal{M}_0|$  in this test
is too small compared to $|\mathcal{M}_X|$ (except for the $B_1$ irrep),
it is of interest to see what happens when $\mathcal{M}_0$ is enlarged while keeping $\mathcal{M}_X$ fixed.
To this end, the bare Hamiltonian projected onto \(\mathcal{M}_X\) is diagonalized to obtain the lowest $|\mathcal{M}_0|$ states for each irrep.
Those determinants with absolute coefficients larger than 0.05 in at least one state are retained, so as to obtain
enlarged model spaces $\mathcal{M}_0^\prime$, see Table  \ref{table:ModelSpaceInfo}.
It can be seen from Table \ref{tab:paspt2h-enlarged-results} that the size-extensivity errors of PASPT2H
are indeed much reduced for the desired 22 states (see Fig. \ref{GSE}), with the mean absolute error being only 3 meV. Therefore, it is perfectly justified to say
that PASPT2H is near-extensive for the target states.
As for PASPT2X, it turns out that the 104-dimensional closed space $\mathcal{M}_X$ for the $A_1$ irrep is already
beset with the intruder-state problem. Therefore,
we take CAS(4,4) as $\mathcal{M}_X$ for each irrep, just for the proof of concept.
It can be seen from Table \ref{tab:paspt2x-results} that PASPT2X is indeed strictly size-extensive
(NB: the nonzero size-extensivity errors for the $^1A_2$ states are due to spin contaminations).

As a final note, it should be mentioned that some excitations, mostly singles, can have unduely large
uncoupled amplitudes even in PASPT2H, where the energy denominators \eqref{denorminator} are never too small.
Rather, their appearance arises\cite{PASPT2} from the
use of a common set of orbitals to construct reference-specific Fock operators \eqref{Fspin} that
do not satisfy the Brillouin condition.
Such excitations tend to slow down the convergence of the linear equations \eqref{LinearSys}
and can thus be characterized as pseudo-intruder states (PIS). Although eliminating such PISs
does improve the convergence rate\cite{PASPT2}, it may result in additional size-extensivity errors.
It is more appropriate to regularize the PISs in one way or another. This should be combined with
a more robust solver for the linear equations, where full use of the internal structure
of the $\mathbf{A}$ matrix in Eqs. \eqref{A_matrix_elmt_1} and \eqref{A_matrix_elmt_2} should be made.
Work along these directions is being carried out at our laboratory.

\begin{table}
	\centering
	\caption{Model spaces $\mathcal{M}_0$ for equilibrium water  with $M_S = 0$. }
	\begin{threeparttable}
		\begin{tabular}{ccccc}\toprule
	Irrep	&	$A_1$ & $A_2$
	& $B_1$
	& $B_2$ \\\toprule
	oCFG$^a$& $|\mathrm{HF}\rangle$       & $1b_2\rightarrow 2b_1$
	& $3a_1\rightarrow 2b_1$
	& $1b_2\rightarrow 4a_1$         \\
			& $3a_1\rightarrow 4a_1$      & $1b_2\rightarrow 3b_1$
			& $1b_1\rightarrow 4a_1$
			& $3a_11b_2\rightarrow 4a_1^2$   \\
			& $1b_1\rightarrow 2b_1$      & $3a_11b_2\rightarrow 4a_12b_1$
			& $3a_1\rightarrow 3b_1$
			& $1b_2\rightarrow 5a_1$         \\
			& $1b_2^2\rightarrow 4a_1^2$  & $1b_11b_2\rightarrow 4a_1^2$
			& $1b_2^2\rightarrow 4a_12b_1$
			& $1b_11b_2\rightarrow 4a_12b_1$ \\
			& $3a_1\rightarrow 5a_1$      &
			&
			& $3a_11b_2\rightarrow 2b_1^2$   \\
			& $3a_1^2\rightarrow 4a_1^2$  &
			&
			&                                \\
			& $1b_2^2\rightarrow 2b_1^2$  &
			&
			&                                \\ \hline
			$(|M_0|, |M_X|, |M_C|)^b$  & (10, 104, 1234) &  (12, 22, 1234)
			&  (8, 10, 1216)
			& (14, 104, 1216)                    \\
            $(|M_0^\prime|, |M_X|, |M_C|)^c$  & (36, 104, 1234) & (20, 22, 1234)
            & (10, 10, 1216)
            & (40, 104, 1216)    \\ \hline
		\end{tabular}
	\begin{tablenotes}
            \item[a] Orbital configuration.
			\item[b] Sizes of $\mathcal{M}_0$, $\mathcal{M}_X$  and CAS(8,8).
            \item[c] $\mathcal{M}^\prime_0$ is selected from $\mathcal{M}_X$ with absolute coefficients larger than 0.05.
		\end{tablenotes}
\end{threeparttable} \label{table:ModelSpaceInfo}
\end{table}

\begin{table}[htbp]
\small
	\centering
	\caption{Test of generalized size-extensivity of PASPT2M/cc-pVDZ for 22 lowest singlet states
of equilibrium water with model spaces $\mathcal{M}_0$ in Table \ref{table:ModelSpaceInfo}.
		The energies for the ground and excited states are given in \(E_h\) and eV, respectively. $\Delta^{\mathrm{GSE}}$ are given in meV.
 }
	\label{tab:paspt2m-results}
	\begin{tabular}{llrrrrr}
		\toprule
		\multicolumn{1}{c}{State} & \multicolumn{1}{c}{major component} & \multicolumn{1}{c}{\(E_{\mathrm{R/S}}\)} & \multicolumn{1}{c}{\(E_{\mathrm{R}/\mathrm{S}_0}\)} & \multicolumn{1}{c}{\(E_{\mathrm{R}_0/\mathrm{S}}\)} & \multicolumn{1}{c}{\(E_{\mathrm{R/S}}^0\)} & \multicolumn{1}{c}{\(\Delta^{\text{GSE}}\)} \\
		\midrule
		\multirow{7}{*}{\(^1A_1\)}
		& Hartree Fock                 & -58.58902257 & -54.68381307 & -41.92207362 & -38.01686413 & 0.000 \\
		& \(3a_1\rightarrow 4a_1\)     & 11.627 & 11.627 & 20.567 & 20.568 & 0.000 \\
		& \(1b_1\rightarrow 2b_1\)     & 18.875 & 26.205 & 18.877 & 26.207 & 0.000 \\
		& \(1b_2^2\rightarrow 4a_1^2\) & 61.537 & 44.874 & 53.743 & 37.079 & 0.000 \\
		& \(3a_1\rightarrow 5a_1\)     & 27.624 & 27.625 & 38.992 & 38.992 & 0.000 \\
		& \(3a_1^2\rightarrow 4a_1^2\) & 30.708 & 30.708 & 41.135 & 41.136 & 0.000 \\
		& \(1b_2^2\rightarrow 2b_1^2\) & 28.813 & 40.910 & 28.815 & 40.911 & 0.000 \\
		\midrule
		\multirow{5}{*}{\(^1A_2\)}
		& \(1b_2\rightarrow 2b_1\)               & 10.534 & 20.454 & 10.536 & 20.456 & 0.000 \\
		& \(1b_2\rightarrow 3b_1\)               & 22.091 & 35.139 & 22.093 & 35.141 & 0.000 \\
		& \(1b_1 1b_2\rightarrow 4a_1^2\)        & 65.906 & 50.625 & 58.112 & 42.831 & 0.000 \\
		& \(3a_1 1b_2\rightarrow 4a_1 2b_1\)     & 20.690 & 30.941 & 30.772 & 41.023 & 0.000 \\
		& \(3a_1 1b_2\rightarrow 4a_1 2b_1\)     & 22.203 & 32.097 & 31.129 & 41.023 & 0.000 \\
		\midrule
		\multirow{4}{*}{\(^1B_1\)}
		& \(3a_1\rightarrow 2b_1\)              & 21.630 & 21.726 & 22.387 & 22.484 & 0.000 \\
		& \(1b_1\rightarrow 4a_1\)              & 23.733 & 24.204 & 23.820 & 24.291 & 0.000 \\
		& \(3a_1\rightarrow 3b_1\)              & 35.878 & 36.411 & 36.636 & 37.169 & 0.000 \\
		& \(1b_2^2\rightarrow 4a_1 2b_1\)       & 36.511 & 38.908 & 36.599 & 38.995 & 0.000 \\
		\midrule
		\multirow{6}{*}{\(^1B_2\)}
		& \(1b_2\rightarrow 4a_1\)              & 17.127 & 18.452 & 17.214 & 18.540 & 0.000 \\
		& \(3a_1 1b_2\rightarrow 4a_1^2\)       & 27.290 & 28.640 & 37.758 & 39.107 & 0.000 \\
		& \(1b_2\rightarrow 5a_1\)              & 35.209 & 36.558 & 35.614 & 36.964 & 0.000 \\
		& \(3a_1 1b_2\rightarrow 2b_1^2\)       & 31.501 & 42.182 & 32.258 & 42.940 & 0.000 \\
		& \(1b_1 1b_2\rightarrow 4a_1 2b_1\)     & 39.317 & 44.660 & 39.404 & 44.747 & 0.000 \\
		& \(1b_1 1b_2\rightarrow 4a_1 2b_1\)     & 42.212 & 44.660 & 42.300 & 44.747 & 0.000 \\
		\bottomrule
	\end{tabular}
\end{table}

\begin{table}[htbp]
\small
	\centering
	\caption{Test of generalized size-extensivity of PASPT2H/cc-pVDZ for 22 lowest singlet states
of equilibrium water with model spaces $\mathcal{M}_0$ in Table \ref{table:ModelSpaceInfo}.
The energies for the ground and excited states are given in \(E_h\) and eV, respectively. $\Delta^{\mathrm{GSE}}$ are given in meV
(mean absolute error: 21.090 meV).
		 }
	\label{tab:paspt2h-results}
	\begin{tabular}{llrrrrr}
		\toprule
		\multicolumn{1}{c}{State} & \multicolumn{1}{c}{major component} & \multicolumn{1}{c}{\(E_{\mathrm{R/S}}\)} & \multicolumn{1}{c}{\(E_{\mathrm{R}/\mathrm{S}_0}\)} & \multicolumn{1}{c}{\(E_{\mathrm{R}_0/\mathrm{S}}\)} & \multicolumn{1}{c}{\(E_{\mathrm{R/S}}^0\)} & \multicolumn{1}{c}{\(\Delta^{\text{GSE}}\)} \\
		\midrule
		\(^1A_1\) & Hartree Fock & -58.58963726 & -54.68427014 & -41.92227968 & -38.01686413 & 1.318 \\
		\(^1A_2\) & \(1b_2\rightarrow 2b_1\) & 10.482 & 20.465 & 10.480 & 20.456 & -8.195 \\
		\(^1A_1\) & \(3a_1\rightarrow 4a_1\) & 11.476 & 11.474 & 20.567 & 20.568 & 3.106 \\
		\(^1B_2\) & \(1b_2\rightarrow 4a_1\) & 16.940 & 18.459 & 17.021 & 18.540 & -0.244 \\
		\(^1A_1\) & \(1b_1\rightarrow 2b_1\) & 18.602 & 26.208 & 18.587 & 26.207 & 14.341 \\
		\(^1A_2\) & \(3a_1 1b_2\rightarrow 4a_1 2b_1\) & 20.650 & 31.036 & 30.640 & 41.023 & -3.071 \\
		\(^1B_1\) & \(3a_1\rightarrow 2b_1\) & 21.646 & 21.739 & 22.392 & 22.484 & -1.318 \\
		\(^1A_2\) & \(3a_1 1b_2\rightarrow 4a_1 2b_1\) & 22.073 & 32.005 & 31.051 & 41.023 & 40.829 \\
		\(^1A_2\) & \(1b_2\rightarrow 3b_1\) & 22.446 & 35.150 & 22.445 & 35.141 & -8.168 \\
		\(^1B_1\) & \(1b_1\rightarrow 4a_1\) & 23.748 & 24.216 & 23.824 & 24.291 & -1.057 \\
		\(^1B_2\) & \(3a_1 1b_2\rightarrow 4a_1^2\) & 27.357 & 28.599 & 37.822 & 39.107 & 43.969 \\
		\(^1A_1\) & \(3a_1\rightarrow 5a_1\) & 27.665 & 27.657 & 38.991 & 38.992 & 8.890 \\
		\(^1A_1\) & \(1b_2^2\rightarrow 2b_1^2\) & 28.797 & 40.912 & 28.790 & 40.911 & 6.777 \\
		\(^1A_1\) & \(3a_1^2\rightarrow 4a_1^2\) & 30.852 & 30.791 & 41.135 & 41.136 & 61.867 \\
		\(^1B_2\) & \(3a_1 1b_2\rightarrow 2b_1^2\) & 31.498 & 42.157 & 32.264 & 42.940 & 17.125 \\
		\(^1B_2\) & \(1b_2\rightarrow 5a_1\) & 35.305 & 36.564 & 35.678 & 36.964 & 26.214 \\
		\(^1B_1\) & \(3a_1\rightarrow 3b_1\) & 35.896 & 36.424 & 36.642 & 37.169 & -1.318 \\
		\(^1B_1\) & \(1b_2^2\rightarrow 4a_1 2b_1\) & 36.675 & 38.920 & 36.657 & 38.995 & 92.513 \\
		\(^1B_2\) & \(1b_1 1b_2\rightarrow 4a_1 2b_1\) & 39.617 & 44.666 & 39.694 & 44.747 & 4.071 \\
		\(^1B_2\) & \(1b_1 1b_2\rightarrow 4a_1 2b_1\) & 42.607 & 44.666 & 42.648 & 44.747 & 39.740 \\
		\(^1A_1\) & \(1b_2^2\rightarrow 4a_1^2\) & 62.221 & 44.869 & 54.365 & 37.079 & 66.053 \\
		\(^1A_2\) & \(1b_1 1b_2\rightarrow 4a_1^2\) & 65.923 & 50.638 & 58.130 & 42.831 & -13.788 \\
		\bottomrule
	\end{tabular}
\end{table}

\begin{table}[htbp]
\small
	\centering
	\caption{Test of generalized size-extensivity of PASPT2H/cc-pVDZ for 22 lowest singlet states
of equilibrium water with enlarged model spaces $\mathcal{M}_0^\prime$ in Table \ref{table:ModelSpaceInfo}.
The energies for the ground and excited states are given in \(E_h\) and eV, respectively. $\Delta^{\mathrm{GSE}}$ are given in meV
(mean absolute error: 3.044 meV). }
	\label{tab:paspt2h-enlarged-results}
	\begin{tabular}{llrrrrr}
		\toprule
		\multicolumn{1}{c}{State} & \multicolumn{1}{c}{major component} & \multicolumn{1}{c}{\(E_{\mathrm{R/S}}\)} & \multicolumn{1}{c}{\(E_{\mathrm{R}/\mathrm{S}_0}\)} & \multicolumn{1}{c}{\(E_{\mathrm{R}_0/\mathrm{S}}\)} & \multicolumn{1}{c}{\(E_{\mathrm{R/S}}^0\)} & \multicolumn{1}{c}{\(\Delta^{\text{GSE}}\)} \\
		\midrule
\(^1A_1\) & Hartree Fock & -58.59036766 & -54.68456607 & -41.92266667 & -38.01686413 & 0.95 \\
\(^1A_2\) & \(1b_2\rightarrow 2b_1\) & 10.498 & 20.470 & 10.490 & 20.456 & -6.436 \\
\(^1A_1\) & \(3a_1\rightarrow 4a_1\) & 11.461 & 11.448 & 20.579 & 20.568 & 1.133 \\
\(^1B_2\) & \(1b_2\rightarrow 4a_1\) & 16.994 & 18.463 & 17.070 & 18.540 & 0.675 \\
\(^1A_1\) & \(1b_1\rightarrow 2b_1\) & 18.422 & 26.215 & 18.413 & 26.207 & 1.035 \\
\(^1A_2\) & \(3a_1 1b_2\rightarrow 4a_1 2b_1\) & 20.673 & 31.056 & 30.648 & 41.023 & -6.852 \\
\(^1B_1\) & \(3a_1\rightarrow 2b_1\) & 21.666 & 21.747 & 22.403 & 22.484 & -0.026 \\
\(^1A_2\) & \(3a_1 1b_2\rightarrow 4a_1 2b_1\) & 21.965 & 31.934 & 31.058 & 41.023 & -4.322 \\
\(^1A_2\) & \(1b_2\rightarrow 3b_1\) & 22.466 & 35.158 & 22.454 & 35.141 & -5.973 \\
\(^1B_1\) & \(1b_1\rightarrow 4a_1\) & 23.766 & 24.224 & 23.833 & 24.291 & -0.026 \\
\(^1B_2\) & \(3a_1 1b_2\rightarrow 4a_1^2\) & 27.055 & 28.557 & 37.599 & 39.107 & 5.815 \\
\(^1A_1\) & \(3a_1\rightarrow 5a_1\) & 27.568 & 27.555 & 39.003 & 38.992 & 1.210 \\
\(^1A_1\) & \(1b_2^2\rightarrow 2b_1^2\) & 28.778 & 40.920 & 28.764 & 40.911 & 5.284 \\
\(^1A_1\) & \(3a_1^2\rightarrow 4a_1^2\) & 30.651 & 30.638 & 41.147 & 41.136 & 1.964 \\
\(^1B_2\) & \(3a_1 1b_2\rightarrow 2b_1^2\) & 31.451 & 42.116 & 32.274 & 42.940 & -0.026 \\
\(^1B_2\) & \(1b_2\rightarrow 5a_1\) & 35.075 & 36.583 & 35.456 & 36.964 & 0.144 \\
\(^1B_1\) & \(3a_1\rightarrow 3b_1\) & 35.916 & 36.432 & 36.653 & 37.169 & -0.026 \\
\(^1B_1\) & \(1b_2^2\rightarrow 4a_1 2b_1\) & 36.182 & 38.929 & 36.249 & 38.995 & -0.026 \\
\(^1B_2\) & \(1b_1 1b_2\rightarrow 4a_1 2b_1\) & 38.375 & 44.674 & 38.445 & 44.747 & 2.769 \\
\(^1B_2\) & \(1b_1 1b_2\rightarrow 4a_1 2b_1\) & 40.730 & 44.674 & 40.800 & 44.747 & 2.617 \\
\(^1A_1\) & \(1b_2^2\rightarrow 4a_1^2\) & 60.740 & 44.884 & 52.927 & 37.079 & 8.545 \\
\(^1A_2\) & \(1b_1 1b_2\rightarrow 4a_1^2\) & 65.942 & 50.646 & 58.139 & 42.831 & -11.130 \\
		\bottomrule
	\end{tabular}
\end{table}

\begin{table}[htbp]
\small
	\centering
	\caption{Test of generalized size-extensivity of PASPT2X/cc-pVDZ for 20 lowest singlet states of equilibrium water with CAS(4,4) as $\mathcal{M}_X$.
		The energies for the ground and excited states are given in \(E_h\) and eV, respectively. $\Delta^{\mathrm{GSE}}$ are given in meV.
		The $\langle S^2\rangle$ values for the spin-contaminated states are given in parentheses.
	}
	\label{tab:paspt2x-results}
	\begin{tabular}{llllllc}
		\toprule
		\multicolumn{1}{c}{State} & \multicolumn{1}{c}{major component} & \multicolumn{1}{c}{\(E_{\mathrm{R/S}}\)} & \multicolumn{1}{c}{\(E_{\mathrm{R}/\mathrm{S}_0}\)} & \multicolumn{1}{c}{\(E_{\mathrm{R}_0/\mathrm{S}}\)} &
		\multicolumn{1}{c}{\(E_{\mathrm{R/S}}^0\)} & \multicolumn{1}{c}{\(\Delta^{\text{GSE}}\)} \\
		\midrule
		\multirow{8}{*}{\(^1A_1\)}
		& Hartree Fock                & -58.58935495 & -54.68407682 & -41.92214225 & -38.01686412 & 0.000 \\
		& \(3a_1\rightarrow 4a_1\)    & 11.517 & 11.517 & 20.568 & 20.568 & 0.000 \\
		& \(1b_2^2\rightarrow 4a_1^2\)& 61.546 & 44.881 & 53.744 & 37.079 & 0.000 \\
		& \(3a_1^2\rightarrow 4a_1^2\)& 30.861 & 30.861 & 41.136 & 41.136 & 0.000 \\
		& \(1b_2^2\rightarrow 3b_1^2\)& 28.744 & 40.911 & 28.744 & 40.911 & 0.000 \\
		& \(3a_1^2\rightarrow 3b_1^2\)& 71.101 & 63.463 & 52.606 & 44.968 & 0.000 \\
		& \(3a_1 1b_2^2\rightarrow 4a_1 2b_1^2\) & 40.261 & 52.428 & 49.312 & 61.479 & 0.000 \\
		& \(3a_1^2 1b_2^2\rightarrow 4a_1^2 2b_1^2\) & 59.606 & 71.773 & 69.880 & 82.047 & 0.000 \\
		\midrule
		\multirow{4}{*}{\(^1A_2\)}
		& \(1b_2\rightarrow 2b_1\)                 & 10.563 & 20.456 & 10.568 & 20.456 & -4.776 \\
		& \(1b_2 3a_1\rightarrow 4a_1 2b_1\)       & 20.789 (2.00) & 31.039 (1.40) & 30.779 & 41.023 & -4.869 \\
		& \(1b_2 3a_1\rightarrow 4a_1 2b_1\)       & 22.080 & 31.973 (0.55) & 31.135 & 41.023 & -4.776 \\
		& \(3a_1^2 1b_2\rightarrow 4a_1^2 2b_1\)   & 41.424 & 51.317 (0.48) & 51.703 & 61.591 & -4.776 \\
		\midrule
		\multirow{4}{*}{\(^1B_1\)}
		& \(3a_1\rightarrow 2b_1\)                 & 21.611 & 21.706 & 22.389 & 22.484 & 0.000 \\
		& \(1b_2^2\rightarrow 4a_1 2b_1\)          & 36.445 & 38.915 & 36.525 & 38.995 & 0.000 \\
		& \(3a_1^2\rightarrow 4a_1 2b_1\)          & 42.613 & 42.707 & 42.957 & 43.052 & 0.000 \\
		& \(3a_1 1b_2^2\rightarrow 4a_1^2 2b_1\)   & 46.640 & 49.110 & 57.093 & 59.563 & 0.000 \\
		\midrule
		\multirow{4}{*}{\(^1B_2\)}
		& \(1b_2\rightarrow 4a_1\)                 & 17.111 & 18.459 & 17.192 & 18.540 & 0.000 \\
		& \(3a_1 1b_2\rightarrow 4a_1^2\)          & 27.307 & 28.655 & 37.760 & 39.107 & 0.000 \\
		& \(3a_1 1b_2\rightarrow 2b_1^2\)          & 31.456 & 42.135 & 32.260 & 42.940 & 0.000 \\
		& \(3a_1^2 1b_2\rightarrow 4a_1 2b_1^2\)    & 52.480 & 63.160 & 52.828 & 63.507 & 0.000 \\
		\bottomrule
	\end{tabular}
\end{table}

\section{Conclusions}\label{Conclusion}
Three variants of PASPT2 have been formulated. Both PASPT2M and PASPT2X are strictly size-extensive (and size-consistent).
However, both are plagued by the intruder-state problem. In particular, PASPT2M has only limited accuracy. In contrast,
the original variant of PASPT2, PASPT2H, is intruder-free for all target states supported by the chosen PAS
(which can readily be generated by sCI). As confirmed by comparing with PASPT2X and by numerical evidence as well,
PASPT2H is near-extensive and the weak violation of size-extensivity has hardly numerical consequence on the target states. Therefore,
the original claim\cite{PASPT2} that `` PASPT2 is up to date the only size-extensive,
size-consistent and intruder-free PAS-based multi-state MRPT2'' is fully justified.

\section*{Acknowledgments}
This work was supported by the National Natural Science Foundation of China (Grant Nos. 22373057 and 22503051).

\section*{Data availability}
No new data were generated or analyzed in this study.

\section*{Conflicts of interest}
There are no conflicts to declare.

\bibliography{iCI}

\end{document}